\documentclass[english,aps,preprint,showpacs,superscriptaddress]{revtex4}
\usepackage[T1]{fontenc}
\usepackage[latin9]{inputenc}
\usepackage{amsmath}
\usepackage{amssymb}
\usepackage{graphicx}
\usepackage{esint}

\makeatletter

\providecommand{\tabularnewline}{\\}

\@ifundefined{textcolor}{}
{%
 \definecolor{BLACK}{gray}{0}
 \definecolor{WHITE}{gray}{1}
 \definecolor{RED}{rgb}{1,0,0}
 \definecolor{GREEN}{rgb}{0,1,0}
 \definecolor{BLUE}{rgb}{0,0,1}
 \definecolor{CYAN}{cmyk}{1,0,0,0}
 \definecolor{MAGENTA}{cmyk}{0,1,0,0}
 \definecolor{YELLOW}{cmyk}{0,0,1,0}
 }

\usepackage{babel}

\makeatother

\usepackage{babel}
\begin{document}

\title{Gravitational Contributions to Gauge Green's Functions and Asymptotic Free Power-Law Running of Gauge Coupling}

\author{Yong Tang}

\email{ytang@itp.ac.cn}

\affiliation{ Kavli Institute for Theoretical Physics China (KITPC) \\
 State Key Laboratory of Theoretical Physics (SKLTP)\\
 Institute of Theoretical Physics, Chinese Academy of Sciences, Beijing,
100190, China }

\affiliation{Physics Division, National Center for Theoretical Sciences, \\
National Tsing Hua University, Hsinchu}

\author{Yue-Liang Wu}

\email{ylwu@itp.ac.cn}

\affiliation{ Kavli Institute for Theoretical Physics China (KITPC) \\
 State Key Laboratory of Theoretical Physics (SKLTP)\\
 Institute of Theoretical Physics, Chinese Academy of Sciences, Beijing,
100190, China }

\begin{abstract}
We perform an explicit one-loop calculation for the gravitational contributions to the two-, three- and four-point gauge Green's functions with paying attention to the quadratic divergences. It is shown for the first time in the diagrammatic calculation that the Slavnov-Taylor identities are
preserved even if the quantum graviton effects are included at one-loop level, such a conclusion is independent of the choice of regularization schemes. We also present a regularization scheme independent calculation based on the gauge condition independent background field framework of Vilkovisky-DeWitt's effective action with focusing on both the quadratic divergence and quartic divergence that is not discussed before. With the harmonic gauge condition, the results computed by using the traditional background field method can consistently be recovered from the Vilkovisky-DeWitt's effective action approach by simply taking a limiting case, and are found to be the same as the ones yielded by the diagrammatic calculation. As a consequence, in all the calculations, the symmetry-preserving and divergent-behavior-preserving loop regularization method can consistently lead to a nontrivial gravitational contribution to the gauge coupling constant with an asymptotic free power-law running at one loop near the Planck scale.
\end{abstract}


\maketitle

\section{Introduction}

The classical theory of general relativity has been well verified
since its establishment in the beginning of last century. However,
the quantum theory of general relativity remains one of the most interesting
and frustrating questions. From the standard renormalization analysis,
the mass dimension of the coupling $\kappa=\sqrt{32\pi G}$ is negative,
which means that general relativity is not a renormalizable theories
\cite{Hooftgravity,Deser1,Deser2}. Since the quantum effects of gravity
become important only at the Planck scale $G^{1/2}\approx10^{19}$GeV,
it may suggest that we can treat it an effective field theory \cite{Donoghue-1,Burgess}
at low energy scales.

Gravitational contribution to gauge theories has attracted much attention
in recent years. Robinson and Wilczek \cite{RW} calculated gravitational
corrections to gauge theories in the framework of traditional background-field
method, and showed that these corrections can render all gauge theories
asymptotically free by changing the gauge couplings to power-law running.
This calculation was done in a specific gauge and cut-off regularization.
However, it was showed in \cite{Pietrykowski} that the result obtained in
\cite{RW} was gauge condition dependent, and the gravitational correction
to $\beta$ function at one-loop order was absent in the harmonic
gauge. Also, it was found in \cite{Toms} that, by using gauge-condition independent
formalism \cite{Vilkovisky,DeWitt}, the gravitational corrections
to the $\beta$ function vanished in dimensional regularization \cite{DimR}.
The above calculations were only involved with gauge two-point Green's
function. Later, the authors in \cite{Ebert} performed a diagrammatic
calculation of two- and three-point Green's functions in the harmonic
gauge by using both cut-off and dimensional regularization schemes,
the same conclusion was yielded that quadratic divergences are absent. 
We should note that all the conclusions are based on one-loop calculations at low energy scale. At or above the Planck scale, the above approximation may break down and new framework for quantum gravity is needed. In this paper, we limit our discussion at one-loop level.

In ref. \cite{TangWu}, we have checked all the calculations in the
framework of diagrammatic and traditional background field methods,
and demonstrated that the results are not only gauge condition dependent
but also regularization scheme dependent. A new consistent loop regularization(LORE)
method \cite{YLW} has been applied to carry out the same calculations
\cite{TangWu} by using both the diagrammatic and traditional background-field
methods. As a consequence, it was found in \cite{TangWu} that there
is asymptotic freedom with power-law running in the harmonic gauge
condition. Further, various approaches were used to discuss similar issues \cite{Reuter,WuFeng,Rodigast,Zanusso,Mackey,Gerwick:2010kq,Folkerts:2011jz, Donoghue, Ellis, TangWu2010}.

In this paper, we shall use both diagrammatic approach and Vilkovisky-DeWitt's effective action to calculate in detail the one-loop gravitational corrections to gauge Green's functions and demonstrate explicitly how the gauge invariance is preserved by these corrections. In diagrammatic calculation, two-, three- and four-point gauge Green's functions are computed in a general way. We will show that the Slavnov-Taylor identities are satisfied irrespective of the regularization schemes. Meanwhile, we will also present a calculation by adopting the Vilkovisky-DeWitt's formalism in Einstein-Maxwell system. Both quadratic and quartic divergences can appear in the one loop corrections and thus a proper regularization scheme needs to be applied to handle the quartic divergences to maintain the gauge invariance.

The paper is organized as follows. In Sec. \ref{sec:Diagrammatic}, we carry out a detailed calculation of one loop gravitational contributions to  two-, three- and four-point gauge Green's functions. As a byproduct, the gravitational contribution to the $\beta$ function of gauge coupling is obtained. In Sec.\ref{sec:Einstein-maxwell}, we apply the Vilkovisky-DeWitt's formalism to the Einstein-Maxwell
system and show the necessary pieces to calculate the gravitational corrections to the $\beta$ function of gauge coupling. In Sec. \ref{sec:Quadratic}, it is shown that the quadratic divergences are presented in a general way, the effects from different regularization schemes are analyzed. Then in Sec. \ref{sec:Quartic}, we focus the discussion on the quartic divergence which in general violates gauge invariance and requires proper regularization schemes to handle it. In the end, we shall summarize our results.

\section{Diagrammatic Calculation}\label{sec:Diagrammatic}

\subsection{Formalism}

The interest of this section is based on the action of Einstein-Yang-Mills theory,
\begin{equation}
\textrm{S}=\int d^{4}x\mathcal{L}=\int d^{4}x\sqrt{-g}\left[\frac{2}{\kappa^{2}}R-\frac{1}{4}g^{\mu\alpha}g^{\nu\beta}\mathcal{F}_{\mu\nu}^{a}\mathcal{F}_{\alpha\beta}^{a}\right],\label{LEYM}
\end{equation}
where $R$ is Ricci scalar, $\mathcal{F}_{\mu\nu}^{a}$ is Yang-Mills
fields strength $\mathcal{F}_{\mu\nu}^{a}=\partial_{\mu}\mathcal{A}_{\nu}^{a}-\partial_{\nu}\mathcal{A}_{\mu}^{a}-ig_{0}[\mathcal{A}_{\mu},\mathcal{A}_{\nu}]$
and $\kappa=\sqrt{32\pi\textrm{G}}$. Here and after, repeated indices
are summed over in the Einstein summation convention. We expand the
metric tensor around a background metric $\bar{g}_{\mu\nu}$ and treat
graviton field as quantum fluctuation $h_{\mu\nu}$ propagating on
the background space-time determined by $\bar{g}_{\mu\nu}$,
\begin{equation}
g_{\mu\nu}=\bar{g}_{\mu\nu}+\kappa h_{\mu\nu}.\label{eq:MetricExpansion}
\end{equation}
 Due to the negative mass dimension of coupling constant $\kappa=\sqrt{16\pi\textrm{G}}$,
this theory is not renormalizable.

The above expansion eq.~(\ref{eq:MetricExpansion}) is exact, but
the expansions of inverse metric and determinant are approximate by
ignoring higher-order terms in realistic calculation. To the second
order in $\kappa$, we have
\begin{eqnarray}
 &  & g^{\mu\nu}=\bar{g}^{\mu\nu}-\kappa h^{\mu\nu}+\kappa^{2}h^{\mu}{}_{\alpha}h^{\alpha\nu},\nonumber \\
 &  & \sqrt{-g}=\sqrt{-\bar{g}}\left[1+\frac{1}{2}\kappa h-\frac{1}{4}\kappa^{2}\left(h^{\mu\nu}h_{\mu\nu}-\frac{1}{2}h^{2}\right)\right].
\end{eqnarray}
 The above expansions are two infinite series and the truncation is
up to the question considered. We only have to keep terms of order
$\kappa$ or $\kappa^{2}$ when considering the gravitational one-loop
correction to pure gauge Green's functions without external graviton
line.

For simplicity, we shall consider the case with flat background space-time,
$\bar{g}_{\mu\nu}=\eta_{\mu\nu}$, where $\eta_{\mu\nu}$ is the Minkowski
metric, $(1,-1,-1,-1)$. The lagrangian can be arranged to different
orders of $h_{\mu\nu}$ or $\kappa$. In the gravity part, we work
with the de Donder harmonic gauge
\[
C^{\mu}=\partial_{\nu}h^{\mu\nu}-\frac{1}{2}\partial^{\mu}h_{\nu}^{\nu}=0,
\]
then, the quadratic terms of $h_{\mu\nu}$ in lagrangian give the graviton's propagator,
\begin{equation}
P_{G}^{\mu\nu\rho\sigma}(k)=\frac{i}{2k^{2}}\left[\eta^{\nu\rho}\eta^{\mu\sigma}+\eta^{\mu\rho}\eta^{\nu\sigma}-\eta^{\mu\nu}\eta^{\rho\sigma}\right].
\end{equation}
 Graviton shall be labelled as double wiggly line in the Feynman diagrams.
For the gauge part, Feynman gauge is used. The interactions of gauge
field and gravity field are determined by expanding the second term
of the lagrangian (\ref{LEYM}). And various vertex functions could
be derived \cite{Ebert}.

\subsection{Renormalization}

In Minkowski space-time, the lagrangian for pure Yang-Mills theory is
\begin{align*}
\mathcal{L}= & -\frac{1}{4}\mathcal{F}_{\mu\nu}^{a}\mathcal{F}^{a\mu\nu}=-\frac{1}{4}\left[\partial_{\mu}\mathcal{A}_{\nu}^{a}-\partial_{\nu}\mathcal{A}_{\mu}^{a}\right]^{2}\\
 & -g_{0}f_{abc}\left(\partial_{\mu}\mathcal{A}_{\nu}^{a}\right)\mathcal{A}^{b\mu}\mathcal{A}^{c\nu}-\frac{1}{4}g_{0}^{2}\left(f_{abe}\mathcal{A}_{\mu}^{a}\mathcal{A}_{\nu}^{b}\right)\left(f_{cde}\mathcal{A}^{c\mu}\mathcal{A}^{d\nu}\right),
\end{align*}
$\mathcal{A}_{\mu}^{a}$ and $g_{0}$ in the above lagrangian are
bare quantities. To remove the divergences appearing in perturbative
calculations, both $\mathcal{A}_{\mu}^{a}$ and $g_{0}$ need to be renormalized,
\[
\mathcal{A}_{\mu}^{a}=z_{2}^{1/2}A_{\mu}^{a},\; g_{0}=z_{g}g,
\]
 $z_{2}$ and $z_{g}$ are referred as field and coupling renormalization
constant, respectively. One can also incorporate renormalization into
the three and four-point vertices as follows,
\begin{align}
\mathcal{L}= & -\frac{1}{4}z_{2}\left[\partial_{\mu}A_{\nu}^{a}-\partial_{\nu}A_{\mu}^{a}\right]^{2}-z_{3}gf_{abc}\left(\partial_{\mu}A_{\nu}^{a}\right)A^{b\mu}A^{c\nu}-\frac{1}{4}z_{4}g^{2}\left(f_{abe}A_{\mu}^{a}A_{\nu}^{b}\right)\left(f_{cde}A^{c\mu}A^{d\nu}\right)\nonumber \\
= & -\frac{1}{4}\left[\partial_{\mu}A_{\nu}^{a}-\partial_{\nu}A_{\mu}^{a}\right]^{2}-gf_{abc}\left(\partial_{\mu}A_{\nu}^{a}\right)A^{b\mu}A^{c\nu}-\frac{1}{4}g^{2}\left(f_{abe}A_{\mu}^{a}A_{\nu}^{b}\right)\left(f_{cde}A^{c\mu}A^{d\nu}\right)\nonumber \\
 & -\frac{1}{4}\delta_{2}\left[\partial_{\mu}A_{\nu}^{a}-\partial_{\nu}A_{\mu}^{a}\right]^{2}-\delta_{3}gf_{abc}\left(\partial_{\mu}A_{\nu}^{a}\right)A^{b\mu}A^{c\nu}-\frac{1}{4}\delta_{4}g^{2}\left(f_{abe}A_{\mu}^{a}A_{\nu}^{b}\right)\left(f_{cde}A^{c\mu}A^{d\nu}\right)\label{eq:counterlagrangian}
\end{align}
 with counterterms
\[
\delta_{2}=z_{2}-1,\;\delta_{3}=z_{3}-1,\;\delta_{4}=z_{4}-1.
\]

The renormalization constants $z_{3}$ and $z_{4}$ are determined
by the divergent part of three- and four-point gauge Green's functions.
And both of them have connections with $z_{g}$ due to gauge invariance,
which is well-known as Slavnov-Taylor \cite{SlavnovTaylor} or Ward identities,
\begin{equation}
z_{g}=\frac{z_{3}}{z_{2}^{3/2}}=\frac{z_{4}^{1/2}}{z_{2}}.\label{eq:Slavnov-Taylor}
\end{equation}
 When fermions and ghosts come in, similar relations exist for their
renormalization constants. The running of gauge coupling with renormalization
scale $\mu$ is described by $\beta$ function whose definition is
\[
\beta(g)\equiv\mu\frac{\partial}{\partial\mu}g.
\]
 With eq.~\ref{eq:Slavnov-Taylor}, one can easily have
\begin{equation}
\beta(g)=g\mu\frac{\partial}{\partial\mu}(\frac{3}{2}\delta_{2}-\delta_{3})=g\mu\frac{\partial}{\partial\mu}(\delta_{2}-\frac{1}{2}\delta_{4}).\label{eq:Beta-function}
\end{equation}

When we consider the system described by eq.~(\ref{LEYM}), with expanding the metric
as eq.~(\ref{eq:MetricExpansion}), many unrenormalizable interactions
come in. Even if we only evaluate one loop gravitational corrections
to gauge Green's function, operators of higher mass dimension, such
as $D_{\rho}F_{\mu\nu}^{a}D^{\rho}F^{a\mu\nu}$, need to be enclosed
in the lagrangian. In this paper, we shall limit our discussion in
the gravitational contributions to operators appearing in eq.~(\ref{eq:counterlagrangian}).
We label the contributions from graviton with a superscript $\kappa$,
\begin{equation}
\beta^{\kappa}_g=g\mu\frac{\partial}{\partial\mu}(\frac{3}{2}\delta_{2}^{\kappa}-\delta_{3}^{\kappa})=g\mu\frac{\partial}{\partial\mu}(\delta_{2}^{\kappa}-\frac{1}{2}\delta_{4}^{\kappa}).\label{grbeta}
\end{equation}

Since the interactions of gauge boson and graviton are gauge invariant,
the Slavnov-Taylor identities should be preserved automatically. The
preservation actually is not trivial at least for two reasons. Firstly,
in the realistic calculation, a gauge condition has to be chosen as a gauge fixing condition which generally spoils
the gauge invariance, which could potentially destroy Slavnov-Taylor
identities. Secondly, at one or higher loop orders, divergences appearing
in the loop momentum integral can also break the identities if an
improper regularization scheme is used. We shall show explicitly that
Slavnov-Taylor identities is maintained and irrespective of the regularization
schemes as well.

\subsection{Diagrammatical Calculation}

In this subsection, we are going to calculate the quadratic divergences
of two, three and four point Green's functions of gauge field. As
a byproduct, we can get the $\beta$ function for the gauge coupling
constant. At first, the counterterms in the last line of eq.~(\ref{eq:counterlagrangian})
give vertex functions,
\begin{eqnarray}
\delta\Pi_{ab}^{\mu\nu} & = & i\delta_{ab}Q^{\mu\nu}\delta_{2},\;\delta T_{abc}^{\mu\nu\rho}(p,q,k)=gf_{abc}V_{pqk}^{\mu\nu\rho}\delta_{3},\;\delta T_{abcd}^{\mu\nu\rho\sigma}=-ig^{2}F_{abcd}^{\mu\nu\rho\sigma}\delta_{4},\\
Q^{\mu\nu} & \equiv & q^{\mu}q^{\nu}-q^{2}\eta^{\mu\nu},\; V_{pqk}^{\mu\nu\rho}\equiv\eta^{\mu\nu}(p-q)^{\rho}+\eta^{\nu\rho}(q-k)^{\mu}+\eta^{\rho\mu}(k-p)^{\nu},\nonumber \\
F_{abcd}^{\mu\nu\rho\sigma} & \equiv & f_{abe}f_{cde}\left(\eta^{\mu\rho}\eta^{\nu\sigma}-\eta^{\mu\sigma}\eta^{\nu\rho}\right)+(b,\nu)\leftrightarrow(c,\rho)+(b,\nu)\leftrightarrow(d,\sigma).
\end{eqnarray}

In gauge theories without gravity, the counter-terms are logarithmically
divergent as the quadratic divergences cancel each other(with proper
regularization schemes used) due to gauge symmetry. However, if gravitational
corrections are taken into account, divergent behavior becomes different.
On dimensional ground, it is known that quadratic divergences can
appear and will contribute to the counter-terms defined above, so
that they will also lead to the corrections to the $\beta$ function.
In later calculations, we will omit the logarithmic divergences and
only focus on the quadratic divergences.

\begin{figure}
\includegraphics[scale=0.4]{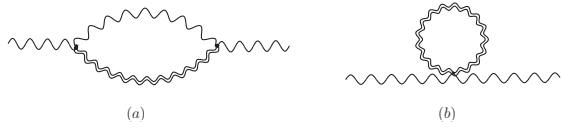} \caption{Graviton loop correction to the gauge two-point Green's function.}
\label{fig:2Gdiag}
\end{figure}

\begin{figure}
\includegraphics[scale=0.4]{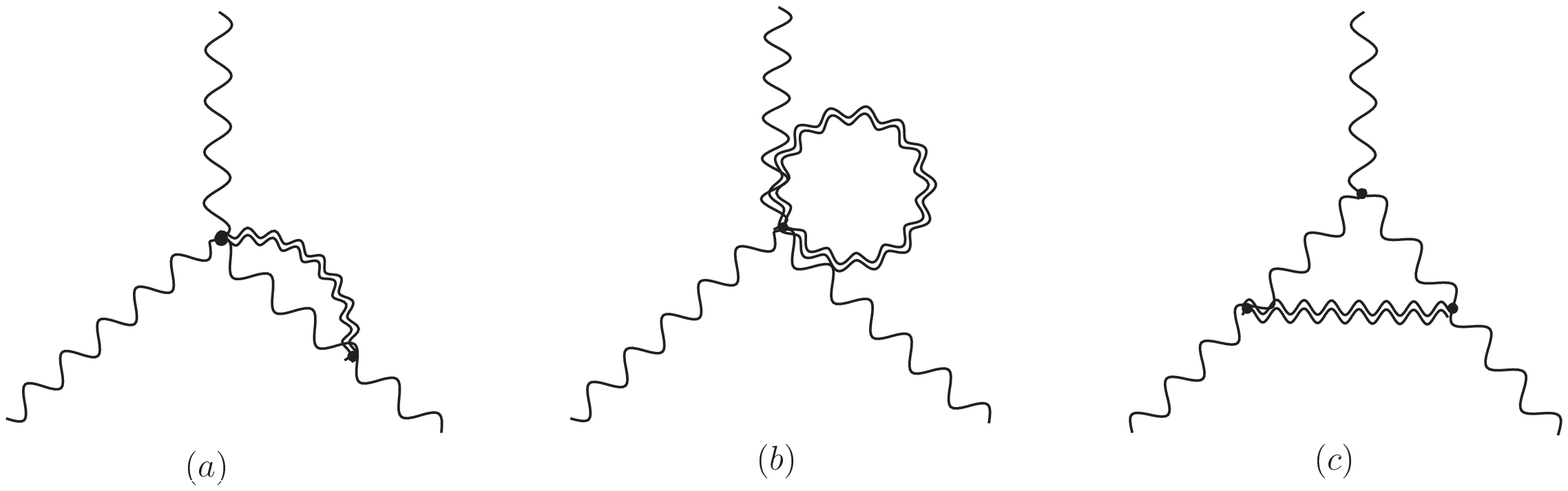} \caption{Graviton loop correction to the gauge three-point Green's function.}
\label{fig:3Gdiag}
\end{figure}

It can be shown that at one-loop level, gravity will contribute to
two- and three-point Green's functions as in the following Feynman
diagrams, Figs.~\ref{fig:2Gdiag} and \ref{fig:3Gdiag}.
For four-point Green's function, two more vertex functions need to
be considered, four gauge bosons--one graviton vertex
\begin{equation}
\mathcal{L}_{{\rm {4Y1G}}}=-\frac{1}{4}g^{2}\kappa f_{abe}f_{cde}A_{\mu}^{a}A_{\nu}^{b}A_{\rho}^{c}A_{\sigma}^{d}\left(\frac{1}{2}\eta^{\mu\rho}\eta^{\nu\sigma}h-\eta^{\mu\rho}h^{\nu\sigma}-h^{\mu\rho}\eta^{\nu\sigma}\right),
\end{equation}
 and four gauge bosons--two gravitons vertex
\begin{align}
\mathcal{L}_{{\rm {4Y2G}}}= & -\frac{1}{4}g^{2}\kappa^{2}f_{abe}f_{cde}A_{\mu}^{a}A_{\nu}^{b}A_{\rho}^{c}A_{\sigma}^{d}\left(\frac{1}{8}\eta^{\mu\rho}\eta^{\nu\sigma}\left[h^{2}-2h^{\alpha\beta}h_{\alpha\beta}\right]\right.\nonumber \\
 & \left.+\eta^{\mu\rho}h^{\nu\beta}h_{\beta}{}^{\sigma}+h^{\mu\alpha}h_{\alpha}{}^{\rho}\eta^{\nu\sigma}-\frac{1}{2}h\left[h^{\mu\rho}\eta^{\nu\sigma}+\eta^{\mu\rho}h^{\nu\sigma}\right]+h^{\mu\rho}h^{\nu\sigma}\right).
\end{align}
 Feynman rules for such vertices can be obtained by standard procedures.
The vertex functions are very complicated with many lorentz indices
and hundreds of terms, and we evaluate the tensor contraction with \textit{FeynCalc}
package \cite{feyncalc}.


To make the results compact, we introduce the tensor type and scalar type irreducible loop integrals(ILIs) at one-loop level,
\begin{equation}
\mathcal{I}_{2}(\mathcal{M}^{2})\equiv\int\frac{d^{4}l}{\left(2\pi\right)^{4}}\frac{1}{l^{2}-\mathcal{M}^{2}},\;\mathcal{I}_{2}^{\mu\nu}(\mathcal{M}^{2})\equiv\int\frac{d^{4}l}{\left(2\pi\right)^{4}}\frac{l^{\mu}l^{\nu}}{\left[l^{2}-\mathcal{M}^{2}\right]^{2}}.\label{eq:ILIs}
\end{equation}
and for short, use $\mathcal{I}_{2}$ and $\mathcal{I}_{2}^{\mu\nu}$
stand for $\mathcal{I}_{2}(0)$ and $\mathcal{I}_{2}^{\mu\nu}(0)$
, respectively.

After tedious calculation, the two-point function gives
\begin{eqnarray*}
\Pi_{ab}^{\mu\nu} & = & \kappa^{2}\delta_{ab}\left[-\frac{1}{2}Q^{\mu\nu}\mathcal{I}_{2}+q^{\mu}q_{\rho}\mathcal{I}_{2}^{\nu\rho}+q^{\nu}q_{\rho}\mathcal{I}_{2}^{\mu\rho}-\eta^{\mu\nu}q_{\rho}q_{\sigma}\mathcal{I}_{2}^{\rho\sigma}-q^{2}\mathcal{I}_{2}^{\mu\nu}\right],
\end{eqnarray*}
and the three point functions from diagrams (a) and (b) of Fig.~\ref{fig:3Gdiag}
are found, when keeping only the quadratically divergent terms, to be
\begin{eqnarray*}
 &  & T_{abc}^{\mu\nu\rho}=\frac{i}{2}g\kappa^{2}f_{abc}\bigg\{\frac{1}{2}V_{qkp}^{\mu\nu\rho}\mathcal{I}_{2}+\left[\eta^{\rho\mu}p_{\sigma}\mathcal{I}_{2}^{\nu\sigma}-\eta^{\mu\nu}p_{\sigma}\mathcal{I}_{2}^{\rho\sigma}+p^{\nu}\mathcal{I}_{2}^{\rho\mu}-p^{\rho}\mathcal{I}_{2}^{\mu\nu}\right]\\
 &  & +\left[\eta^{\mu\nu}q_{\sigma}\mathcal{I}_{2}^{\rho\sigma}-\eta^{\nu\rho}q_{\sigma}\mathcal{I}_{2}^{\mu\sigma}+q^{\rho}\mathcal{I}_{2}^{\mu\nu}-q^{\mu}\mathcal{I}_{2}^{\nu\rho}\right]+\left[\eta^{\nu\rho}k_{\sigma}\mathcal{I}_{2}^{\mu\sigma}-\eta^{\rho\mu}k_{\sigma}\mathcal{I}_{2}^{\nu\sigma}+k^{\mu}\mathcal{I}_{2}^{\nu\rho}-k^{\nu}\mathcal{I}_{2}^{\rho\mu}\right]\bigg\}.
\end{eqnarray*}

\begin{figure}
\includegraphics[scale=0.48]{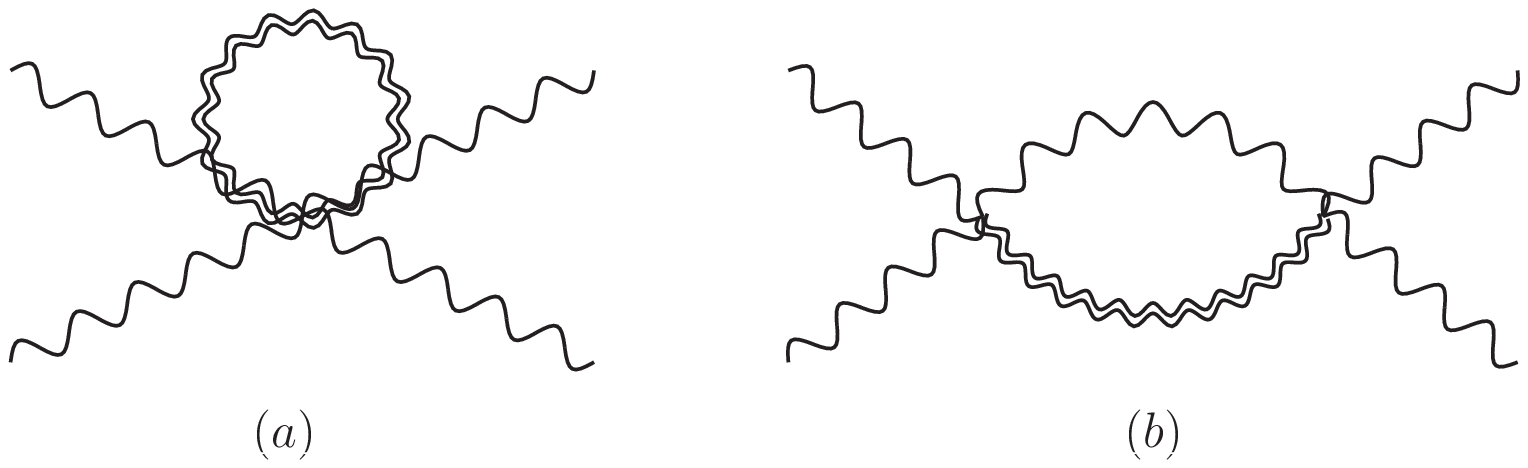}

\includegraphics[scale=0.4]{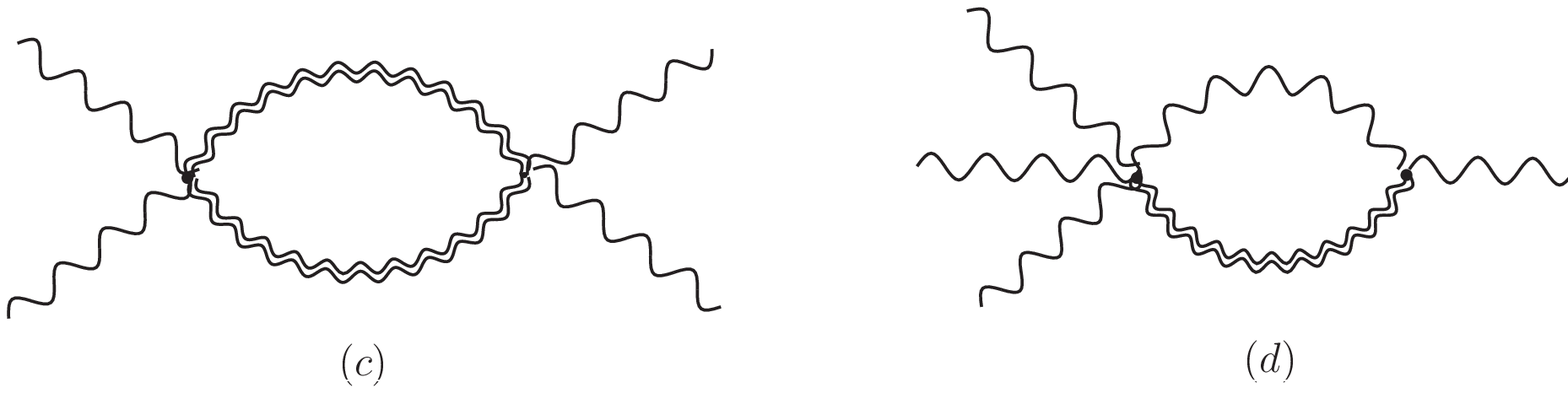}

\includegraphics[scale=0.45]{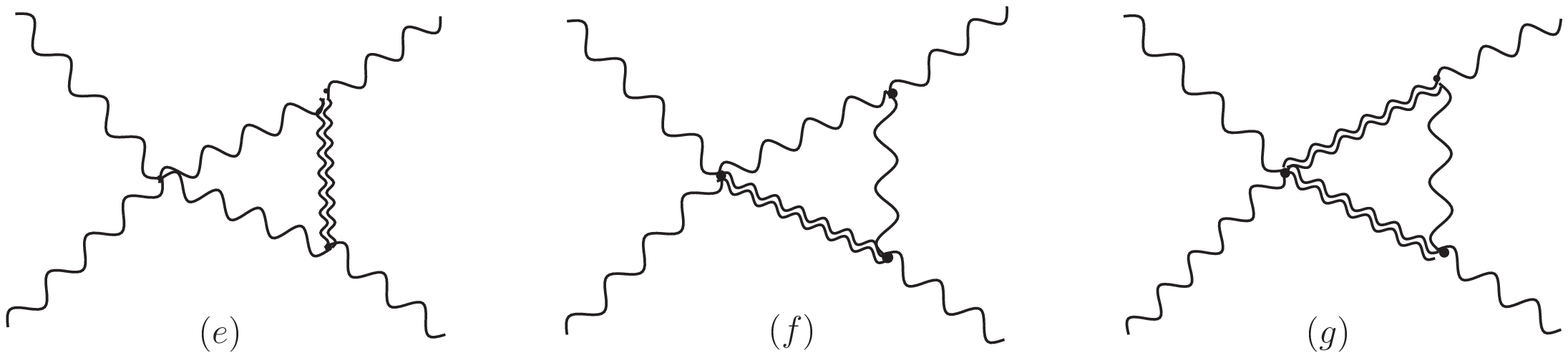}

\includegraphics[scale=0.45]{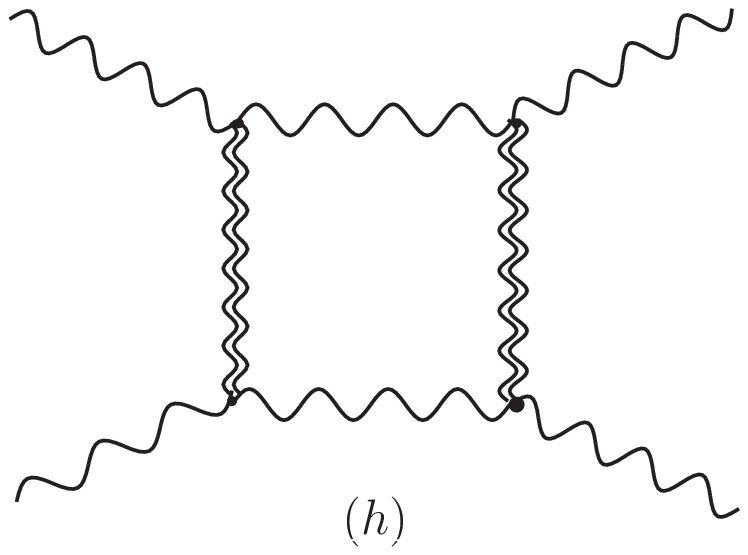} \caption{Graviton loop correction to the gauge four-point Green's function.}
\label{fig:4Gdiag}
\end{figure}

At one loop level, there are a few Feynman diagrams contributing to the four-point Green's function as shown in Fig. \ref{fig:4Gdiag}.
At first sight, the calculation of four point Green's function seems a frustrating task. However, after some analysis on the superficial and real degree of divergence, we can show that only part of these diagrams
have quadratic divergence. Superficial degree of divergence is obtained
by standard renormalization analysis \cite{WeinbergBook}. We summarize it
in the Table I. Where the numbers $2$, $1$ and $0$ stand for quadratic, linear
and logarithmic.

\begin{table}
\begin{tabular}{|c|c|c|c|}
\hline
Diagrams  & Superficial  & Real  & Order\tabularnewline
\hline
(a)  & 2  & 2  & $g^{2}\kappa^{2}$\tabularnewline
\hline
(b)  & 2  & 2  & $g^{2}\kappa^{2}$\tabularnewline
\hline
(c)  & 4  & 0  & $\kappa^{4}$\tabularnewline
\hline
(d)  & 2  & 1  & $g^{2}\kappa^{2}$\tabularnewline
\hline
(e)  & 2  & 0  & $g^{2}\kappa^{2}$\tabularnewline
\hline
(f)  & 2  & 1  & $g^{2}\kappa^{2}$\tabularnewline
\hline
(g)  & 4  & 0  & $\kappa^{4}$\tabularnewline
\hline
(h)  & 4  & 0  & $\kappa^{4}$\tabularnewline
\hline
\end{tabular}\caption{Degree of divergence for each diagram in Fig.~\ref{fig:4Gdiag}.}
\end{table}

Eventually the one-loop gravitational contributions to four point gauge Green's
function are found to be
\begin{eqnarray*}
T_{abcd}^{\mu\nu\rho\sigma} & = & g^{2}\kappa^{2}\bigg\{ f_{abe}f_{cde}\left[\frac{1}{2}\mathcal{I}_{2}\left(\eta^{\mu\rho}\eta^{\nu\sigma}-\eta^{\mu\sigma}\eta^{\nu\rho}\right)-\mathcal{I}_{2}^{\mu\rho}g^{\nu\sigma}+g^{\mu\sigma}\mathcal{I}_{2}^{\nu\rho}+\mathcal{I}_{2}^{\mu\sigma}g^{\nu\rho}-g^{\mu\rho}\mathcal{I}_{2}^{\nu\sigma}\right]\\
 &  & +(b,\nu)\leftrightarrow(c,\rho)+(b,\nu)\leftrightarrow(d,\sigma)\bigg\}
\end{eqnarray*}

Thus the counterterms $\delta_{2},\;\delta_{3}\;{\rm and}\;\delta_{4}$ are
determined by the quadratically divergent part of $\Pi_{ab}^{\mu\nu},\; T_{abc}^{\mu\nu\rho}\;{\rm and}\; T_{abcd}^{\mu\nu\rho\sigma}$,
respectively.
\begin{align*}
\Pi_{ab}^{\mu\nu}+i\delta_{ab}Q^{\mu\nu}\delta_{2}^{\kappa} & \sim0,\\
T_{abc}^{\mu\nu\rho}+gf_{abc}V_{pqk}^{\mu\nu\rho}\delta_{3}^{\kappa} & \sim0,\\
T_{abcd}^{\mu\nu\rho\sigma}-ig^{2}F_{abcd}^{\mu\nu\rho\sigma}\delta_{4}^{\kappa} & \sim0,
\end{align*}
 from which $\delta_{2}^{\kappa}$,$\delta_{3}^{\kappa}$, and $\delta_{4}^{\kappa}$
are determined, respectively. As there are still tensor type quadratic
divergences appearing in the above expressions, we shall first reduce them into the scalar type ones. While the subtle
can be hidden in such a reducing step, namely it will depend on the regularization schemes which spoil either the symmetry or divergence behavior of original theory, we shall discuss such an issue below in detail.

\textit{Regularization}: In the cut-off regularization which is known to spoil gauge and translational symmetries, one has
\begin{equation}
\mathcal{I}_{2}^R=-\frac{i}{16\pi^{2}}\left[\Lambda^{2}-\mu^{2}\right],\quad \mathcal{I}_{2\mu\nu}^R=\frac{1}{4}g_{\mu\nu}\mathcal{I}_{2}^R. \label{cutoffR}
\end{equation}
where the superscript $R$ denote the regularized ones. Putting these formulas into the divergent two-, three- and four-point Green's functions, we straightforwardly get the following results due to cancelations
\begin{eqnarray*}
\delta_{2}^{\kappa} & \sim & 0,\quad \delta_{3}^{\kappa}\sim 0,\quad \delta_{4}^{\kappa}\sim 0.
\end{eqnarray*}
Note that in gauge theories with or without fermions, the relations like eq.~(\ref{cutoffR}) will destroy gauge invariance in two-point Green's functions. In the dimensional regularization which is known to suppress the quadratic divergences, we have $\mathcal{I}_{2}^R=0$ and
then yield the same results as the ones in the cut-off regularization
\begin{eqnarray*}
\delta_{2}^{\kappa} & \sim & 0,\quad \delta_{3}^{\kappa}\sim 0,\quad \delta_{4}^{\kappa}\sim 0.
\end{eqnarray*}
Note that the vanishes of the above functions in both the cut-off regularization and dimensional regularization have different origins.

We shall adopt the consistent loop regularization(LORE) method\cite{YLW} which preserves both symmetries and divergence behavior of original theories and has extensively been applied to various calculations with consistent results\cite{Dai:2003ip, Ma:2005md, Cui:2008uv}. Recently, the consistency and advantage of the LORE method has further been demonstrated by merging with Bjorken-Drell's analogy between Fynman diagrams and electric circuits and also by explicitly applying to the two-loop regularization and renormalization of $\phi^4$ theory\cite{Huang:2011xh}. In the LORE method, we have the following consistency condition of gauge invariance for the regularized irreducible loop integrals
\begin{equation}\label{LRconsistent}
\mathcal{I}_{2\mu\nu}^R=\frac{1}{2}g_{\mu\nu}\mathcal{I}_{2}^R
\end{equation}
with its explicit form given by
\begin{eqnarray}
\mathcal{I}_2^R&=&\frac{-i}{16\pi^2}\{M_c^2-\mu_s^2- \mu_s^2(ln\frac{M_c^2}{\mu_s^2}-\gamma_w)\}
\end{eqnarray}
where $M_c$ and $\mu_s$ play the roles of the ultraviolet and infrared cut-off energy scales. Thus the divergent counterterms are found to be
\begin{eqnarray}
\delta_{2}^{\kappa} & \sim & \frac{i}{2}\kappa^{2}\mathcal{I}_{2}^R,\quad \delta_{3}^{\kappa}\sim\frac{i}{2}\kappa^{2}\mathcal{I}_{2}^R,\quad \delta_{4}^{\kappa}\sim\frac{i}{2}\kappa^{2}\mathcal{I}_{2}^R.\label{eq:LRdiv}
\end{eqnarray}
which can be shown to satisfy the Slavnov-Taylor identities eq.(\ref{eq:Slavnov-Taylor}). In fact, one can easily check that as long as the consistency condition of gauge invariance eq.(\ref{LRconsistent}) is imposed, the Slavnov-Taylor identities eq.(\ref{eq:Slavnov-Taylor}) are preserved no matter which regularization scheme is used.

\textit{The $\beta$ function}: Putting the leading quadratically divergent
parts of $\delta_{2}^{\kappa}$ and $\delta_{3}^{\kappa}$(or $\delta_{4}^{\kappa}$)
into eq.($\ref{grbeta}$), we obtain the gravitational corrections
to the gauge $\beta$ function
\begin{equation}
\beta^{\kappa}_g=-g\kappa^{2}\frac{\mu^{2}}{32\pi^{2}}\label{eq:Diagrambeta}
\end{equation}
which shows that there are gravitational quadratic corrections to
the gauge $\beta$ function when the LORE method is adopted
to evaluate the quadratic divergent integrals, which is different
from the results yielded by using the cut-off and dimensional regularization
schemes.

Note that for an abelian gauge theory, there are no counterterms for $\delta_{3}^{\kappa}$ and $\delta_{4}^{\kappa}$. Thus in the abelian gauge case, the renormalization constant of gauge coupling $z_g$ is related to that of gauge field $z_2$ with $z_gz_2^{1/2}=1$, the corresponding $\beta$-function correction is given via $\beta^{\kappa}_g =\frac{1}{2}g\mu\frac{\partial}{\partial \mu}\delta_{2}^{\kappa}$, which leads to the same result as eq.~(\ref{eq:Diagrambeta}). Therefore, the gravitational correction to the running of gauge coupling is universal for all gauge theories.

\section{Vilkovisky-DeWitt's background field method\label{sec:Einstein-maxwell}}

In this section, we shall apply Vilkovisky-DeWitt's background field
method to Einstein-Maxwell system other than Einstein-Yang-Mills systerm,
for simplicity. This section is partly overlapped with \cite{DJToms}
about the action expansion and with \cite{heatkernel, he, TangWu2010} about quadratic divergences, but we shall present a complete calculation in a regularization independent way and pay attention to the quartic divergence which has not been discussed before. Details of the calculation are given in the appendix.

For a comparison with the results obtained in \cite{DJToms}, we shall use the same notation. Especially, the DeWitt's
condensed index notation is used throughout below, except for places
where an explicit calculation is given. In the appendix, a short review
of the effective action is given. In a general gauge condition, the resulting Vilkovisky-DeWitt's effective action is given
by
\begin{equation}
\Gamma[\bar{\varphi}]=S[\bar{\varphi}]-\ln\det Q^{\alpha}{}_{\beta}+\frac{1}{2}\ln\det\left(\nabla^{i}\nabla_{j}S[\bar{\varphi}]+\frac{1}{2\Omega}\chi_{\alpha}{}^{,i}\chi^{\alpha}{}_{,j}\right).\label{eq:effectiveactionFull}
\end{equation}
where $\chi_{\alpha}$ is the gauge condition, $Q^{\alpha}{}_{\beta}$ is
the Faddeev-Popov factor, $\bar{\varphi}$ is the background field,
and $\nabla_{i}\nabla_{j}S[\bar{\varphi}]=S_{,ij}[\bar{\varphi}]- \Gamma_{ij}^{k}S_{,k}[\bar{\varphi}]$ with $\Gamma_{ij}^{k}$ being given and explained in the appendix.

\subsection{Action Expansion}

We expand the fields, $\varphi^{i}=(g_{\mu\nu},A_{\mu})$, at the
flat background-fields, $\bar{\varphi}^{i}=(\delta_{\mu\nu},\bar{A}_{\mu})$,
\begin{eqnarray}
g_{\mu\nu} & = & \delta_{\mu\nu}+\kappa h_{\mu\nu};\quad A_{\mu}=\bar{A}_{\mu}+a_{\mu}.
\end{eqnarray}
and label $\eta^{i}=(h_{\mu\nu},a_{\nu})$ as the graviton and photon fields. Due to the complicated $\bar{\Gamma}_{ij}^{k}$
, it is much simpler to work in Landau-DeWitt gauge ($\omega=1$),
$K_{\alpha i}[\bar{\varphi}]\eta^{i}=0$. Explicitly, we have
\begin{eqnarray*}
\chi_{\lambda} & = & \frac{2}{\kappa}(\partial^{\mu}h_{\mu\lambda}-\frac{1}{2}\partial_{\lambda}h)+\omega(\bar{A}_{\lambda}\partial^{\mu}a_{\mu}+a^{\mu}\bar{F}_{\mu\lambda})\\
\chi & = & -\partial^{\mu}a_{\mu}.
\end{eqnarray*}
$K_{\alpha}^{i}[\bar{\varphi}]$ is the generator of gauge transformations.
Details can be referred to the appendix. In this gauge, the resulting Vilkovisky-DeWitt's
effective action is given by
\begin{eqnarray*}
\Gamma[\bar{\varphi}] & = & S[\bar{\varphi}]-\ln\det Q_{\alpha\beta}[\bar{\varphi}]\\
 &  & +\frac{1}{2}\lim_{\Omega\rightarrow0}\ln\det\left(\nabla^{i}\nabla_{j}S[\bar{\varphi}]+\frac{1}{2\Omega}K_{\alpha}^{i}[\bar{\varphi}]K_{j}^{\alpha}[\bar{\varphi}]\right)
\end{eqnarray*}
with $\nabla_{i}\nabla_{j}S[\bar{\varphi}]=S_{,ij}[\bar{\varphi}]-\Gamma_{ij}^{k}S_{,k}[\bar{\varphi}]$,
and the connection $\Gamma_{ij}^{k}$ is determined by $g_{ij}[\varphi]$
which is the metric on the field space. In the resulting effective action, a parameter, $v$, is
introduced for the connection term \cite{DJToms},
\begin{equation}
S_{q}=\frac{1}{2}\eta^{i}\left(S_{,ij}-v\Gamma_{ij}^{k}S_{,k}+\frac{1}{2\Omega}K_{\alpha i}K_{j}^{\alpha}\right)\eta^{j}.\label{Eq:vConnection}
\end{equation}
Note that both $\omega$ and $v$ are not real gauge condition parameters,
and their values are actually fixed in Landau-DeWitt gauge, $\omega=1$,
$v=1$. They are introduced here just for an advantage of comparing
with the traditional background field method in harmonic gauge by simply taking
$\omega=0$, $v=0$. In principle, the Vilkovisky-DeWitt formalism is applicable in any gauge condition
as it has been verified to be gauge condition independent\cite{FradkinTseytlin,BV,Huggins}.
While in a practical calculation, such a formalism becomes much simpler
in Landau-DeWitt gauge at one loop. Therefore, we will impose eventually
the Landau-DeWitt gauge condition: $\omega=1$, $v=1$, $\xi\rightarrow0$
and $\zeta\rightarrow0$ to obtain a gauge condition independent result
as guaranteed by the Vilkovisky-DeWitt formalism. Meanwhile, by taking
$\omega=0$, $v=0$, and $\xi=1/\kappa^{2}$, $\zeta=1/2$, we can
straightforwardly read out the result in the traditional background
field method in harmonic gauge.

By appropriately arranging all the terms in the expanded action, we can express $S_{q}$
in eq.(\ref{Eq:vConnection}) as follows
\begin{equation}
S_{q}=S_{0}+S_{1}+S_{2},
\end{equation}
which is found to be consistent with that in \cite{DJToms} where terms from $\bar{A}_{\lambda}\partial^{\mu}a_{\mu}$ in gauge condition are neglected for evaluation of logarithmic divergences. Later we will show that such terms could lead to quartic divergences and violate gauge invariance. 

The free part can be written as
\begin{eqnarray}
S_{0} & = & \int d^{4}x\left[-\frac{1}{2}h^{\mu\nu}\Box h_{\mu\nu}+\frac{1}{4}h\Box h+\left(\frac{1}{\kappa^{2}\xi}-1\right)\left(\partial^{\mu}h_{\mu\nu}-\frac{1}{2}\partial_{\nu}h\right)^{2}-\Lambda\left(h^{\mu\nu}h_{\mu\nu}-\frac{1}{2}h^{2}\right)\right.\nonumber \\
 &  & \left.{}+\frac{1}{2}a_{\mu}\left(-\delta^{\mu\nu}\Box+\partial^{\mu}\partial^{\nu}\right)a_{\nu}+\frac{1}{4\zeta}\left(\partial^{\mu}a_{\mu}\right)^{2}-\frac{v}{2}\Lambda\delta^{\mu\nu}a_{\mu}a_{\nu}\right]
\end{eqnarray}
with $\Lambda$ the cosmological constant. The interaction terms with linear on graviton $h_{\mu\nu}$ or gauge field
$a_{\mu}$ have the following form
\begin{eqnarray*}
S_{1} & = & \frac{\kappa}{2}\int d^{4}x\left(\bar{F}{}^{\mu\nu}h\partial_{\mu}a_{\nu}-2\bar{F}_{\alpha}{}^{\nu}h^{\mu\alpha}\partial_{\mu}a_{\nu}+2\bar{F}_{\alpha}{}^{\nu}h^{\mu\alpha}\partial_{\nu}a_{\mu}\right)\\
 &  & -\frac{\kappa v}{4}\int d^{4}x\left(\delta_{\sigma}^{\lambda}\delta^{\mu\nu}-2\delta_{\sigma}^{(\mu}\delta^{\nu)\lambda}\right)\partial_{\tau}\bar{F}{}^{\sigma\tau}h_{\mu\nu}a_{\lambda}\\
 &  & +\frac{\omega}{\kappa\xi}\int d^{4}x\left(\partial^{\mu}h_{\mu\nu}-\frac{1}{2}\partial_{\nu}h\right)\left(\bar{A}^{\nu}\partial^{\lambda}a_{\lambda}+a^{\lambda}\bar{F}_{\lambda}{}^{\nu}\right)\\
 & = & \int d^{4}x\left[C_{11}^{\alpha\beta\mu\nu}h_{\alpha\beta}\partial_{\mu}a_{\nu}+C_{12}^{\alpha\beta\mu}h_{\alpha\beta}a_{\mu}+\frac{\omega}{\kappa\xi}C_{13}^{\nu\alpha\beta}\partial_{\nu}h_{\alpha\beta}\partial^{\mu}a_{\mu}\right]=S_{11}+S_{12}+S_{13}
\end{eqnarray*}
and the interaction terms with quadratic on graviton $h_{\mu\nu}$ or gauge
field $a_{\mu}$ are given by
\begin{eqnarray}
S_{2} & = & \frac{\kappa^{2}}{4}\int d^{4}x\bar{F}_{\mu\nu}\bar{F}_{\alpha\beta}\left(2\delta^{\mu\alpha}h_{\lambda}^{\nu}h^{\lambda\beta}+h^{\mu\alpha}h^{\nu\beta}-\delta^{\mu\alpha}hh^{\nu\beta}\right)-\frac{\kappa^{2}}{16}\int d^{4}x\bar{F}^{2}\left(h^{\mu\nu}h_{\mu\nu}-\frac{1}{2}h^{2}\right)\nonumber \\
 &  & +\frac{\kappa^{2}v}{4}\int d^{4}x\biggl(\frac{1}{2}\bar{F}{}^{\lambda}{}_{\gamma}\bar{F}{}^{\sigma\gamma}\delta^{\mu\nu}-\bar{F}{}^{\mu}{}_{\gamma}\bar{F}{}^{\sigma\gamma}\delta^{\nu\lambda}+\left[\frac{1}{4}\delta^{\mu\lambda}\delta^{\sigma\nu}-\frac{1}{8}\delta^{\mu\nu}\delta^{\lambda\sigma}\right]\bar{F}^{2}\biggl)h_{\mu\nu}h_{\lambda\sigma}\nonumber \\
 &  & -\int d^{4}x\left[\frac{\kappa^{2}v}{4}\left(\frac{1}{8}\delta^{\mu\nu}\bar{F}^{2}-\frac{1}{2}\bar{F}{}^{\mu}{}_{\gamma}\bar{F}{}^{\nu\gamma}\right)-\frac{\omega^{2}}{4\xi}\bar{F}{}^{\mu}{}_{\gamma}\bar{F}{}^{\nu\gamma}\right]a_{\mu}a_{\nu}+\frac{\omega^{2}}{4\xi}\int d^{4}x\bar{A}_{\lambda}\bar{A}^{\lambda}\partial^{\mu}a_{\mu}\partial^{\nu}a_{\nu}\nonumber \\
 & = & \int d^{4}x\left[C_{21}^{\alpha\beta\mu\nu}h_{\alpha\beta}h_{\mu\nu}+C_{22}^{\mu\nu}a_{\mu}a_{\nu}+\frac{\omega^{2}}{4\xi}C_{23}\partial^{\mu}a_{\mu}\partial^{\nu}a_{\nu}\right]=S_{21}+S_{22}+S_{23}.
\end{eqnarray}
The tensor coefficients, $C_{11}^{\alpha\beta\mu\nu}$, $C_{12}^{\alpha\beta\mu}$,
$C_{13}^{\nu\alpha\beta}$, $C_{21}^{\alpha\beta\mu\nu}$, $C_{22}^{\mu\nu}$
and $C_{23}$ are functions of $\bar{A}_{\mu}$ and $\delta_{\mu\nu}$,
and they can be read out directly. The graviton and photon propagators
are determined by $S_{0}$. And the terms in $S_{1}$ and $S_{2}$
will be treated as interactions between background fields and quantum
fields. Note that $S_{13}$ and $S_{23}$ are proportional to $\omega$,
and they result from the $\bar{A}_{\lambda}\partial^{\mu}a_{\mu}$ term
in the Landau-DeWitt gauge condition for graviton given in eq.(\ref{eq:GaugeForGraviton}).
We shall show that these two terms do not contribute to the effective
action in the present choice of the gauge condition.

We can write the photon propagator in momentum space as \cite{DJToms}
\[
\left\langle a_{\mu}(x)a_{\nu}(x')\right\rangle =G_{\mu\nu}(x,x')=\int\frac{d^{4}p}{(2\pi)^{4}}e^{ip\cdot(x-x')}G_{\mu\nu}(p),
\]
 and the graviton propagator as
\[
\left\langle h_{\rho\sigma}(x)h_{\lambda\tau}(x')\right\rangle =G_{\rho\sigma\lambda\tau}(x,x')=\int\frac{d^{4}p}{(2\pi)^{4}}e^{ip\cdot(x-x')}G_{\rho\sigma\lambda\tau}(p).
\]
 Using the explicit result for $S_{0}$, we have
\begin{equation}
G_{\mu\nu}(p)=\frac{\delta_{\mu\nu}}{p^{2}-v\Lambda}+(2\zeta-1)\frac{p_{\mu}p_{\nu}}{(p^{2}-v\Lambda)(p^{2}-2\zeta v\Lambda)}
\end{equation}
 and
\begin{equation}
G_{\rho\sigma\lambda\tau}(p)=\frac{\delta_{\rho(\lambda}\delta_{\tau)\sigma}-\frac{1}{2}\delta_{\rho\sigma}\delta_{\lambda\tau}}{\left(p^{2}-2\Lambda\right)}+(\kappa^{2}\xi-1)\frac{\delta_{\rho(\lambda}p_{\tau)}p_{\sigma}+\delta_{\sigma(\lambda}p_{\tau)}p_{\rho}}{\left(p^{2}-2\Lambda\right)\left(p^{2}-2\kappa^{2}\xi\Lambda\right)}
\end{equation}

Now we are ready to check that $S_{13}$ and $S_{23}$ give vanishing
quadratic divergences. In the harmonic gauge, $\omega=0$, it becomes manifest.
In the Landau-DeWitt gauge, $\omega=1$ and $\zeta=0$, the gauge propagator
is
\[
G_{\mu\nu}(p)=\frac{1}{p^{2}}\left[\delta_{\mu\nu}-\frac{p_{\mu}p_{\nu}}{p^{2}}\right]
\]
Since the interactions in $S_{13}$ and $S_{23}$ have a factor of
$\partial^{\mu}a_{\mu}$, then $p^{\mu}G_{\mu\nu}$ and $p^{\mu}p^{\nu}G_{\mu\nu}$
will appear in the tensor contraction, which give vanishing contributions.
However, in a general gauge, we will encounter quartic divergent integrals like  $\bar{A}_{\mu}\bar{A}^{\mu}\int\frac{d^{4}p}{(2\pi)^{2}}1$,
such a quartically divergent term has to be well regularized, otherwise it will
violate the $U(1)$ gauge symmetry due to its contribution to the gauge boson mass. We shall discuss it further next
section.

A similar analysis can be made for the effective action of ghost part. The free part can be written down as
\begin{equation}
S_{GH0}=\int d^{4}x\left[-\frac{2}{\kappa^{2}}\bar{c}^{\lambda}\square c_{\lambda}-\bar{c}\square c\right]\label{eq:GhostSGH0}
\end{equation}
The interaction term with linear on gravity ghost or gauge ghost is given by
\begin{eqnarray}
S_{GH1} & = & \int d^{4}x\biggl\{\omega\bar{c}^{\lambda}\bar{F}_{\mu\lambda}c^{,\mu}+\omega\bar{c}^{\lambda}\bar{A}_{\lambda}\square c+\left[\bar{c}\bar{A}_{\nu,\mu}c^{\nu,\mu}-\bar{c}^{,\mu}\bar{A}_{\mu,\nu}c^{\nu}+\bar{c}\bar{A}_{\nu}\Box c^{\nu}\right]\biggl\}\label{eq:GhostSGH1}
\end{eqnarray}
and the interaction term with quadratic terms on gravity ghost has the form
\begin{equation}
S_{GH2}=\omega\int d^{n}x\biggl\{\bar{c}^{\lambda}\bar{F}_{\lambda\mu}\left[\bar{A}^{\mu}{}_{,\nu}c^{\nu}+\bar{A}_{\nu}c^{\nu,\mu}\right]-\left[\bar{c}^{\lambda}\bar{A}_{\lambda}\bar{A}_{\rho}\square c^{\rho}+\bar{c}^{\lambda}\bar{A}_{\lambda}\bar{A}_{\nu,\rho}c^{\rho,\nu}\right]\biggl\}.\label{eq:GhostSGH2}
\end{equation}
 From the free part $S_{GH0}$, the ghosts' propagators can easily be read off
\begin{eqnarray*}
\langle c_{\mu}(x)\bar{c}_{\nu}(x')\rangle & = & \Delta_{\mu\nu}(x,x')=\int\frac{d^{4}p}{(2\pi)^{4}}e^{ip\cdot(x-x')}\Delta_{\mu\nu}(p),\\
\langle c_{\mu}(x)c_{\nu}(x')\rangle & = & \langle\bar{c}_{\mu}(x)\bar{c}_{\nu}(x')\rangle=0\\
\langle c(x)\bar{c}(x')\rangle & = & \Delta(x,x')=\int\frac{d^{4}p}{(2\pi)^{4}}e^{ip\cdot(x-x')}\Delta(p),\\
\langle c(x)c(x')\rangle & = & \langle\bar{c}(x)\bar{c}(x')\rangle=0
\end{eqnarray*}
 where the propagators in the momentum space are given by
\[
\Delta_{\mu\nu}(p)=\frac{\kappa^{2}}{2}\delta_{\mu\nu}\frac{1}{p^{2}},\quad\Delta(p)=\frac{1}{p^{2}}
\]

\subsection{One-Loop quadratically divergent contribution\label{sec:Quadratic}}

In this subsection, we shall present our results for the quadratically
divergent contributions. As a consistent check, we have reproduced the results for the logarithmic divergent
contributions to the $\beta$ function when the cosmological constant
is included\cite{DJToms}. Here we shall not repeat the similar analysis and only carry out the calculation for the leading quadratic divergences
which are encountered in corrections to $\frac{1}{4}F_{\mu\nu}F^{\mu\nu}$.
The contributions from the gravity-gauge interactions to the effective action
can be written as 
\begin{equation}
\Gamma_{G}=\langle S_{2}\rangle-\frac{1}{2}\langle S_{1}^{2}\rangle,\quad\langle S_{2}\rangle=\langle S_{21}\rangle+\langle S_{22}\rangle
\end{equation}
with the explicit forms given by
\begin{eqnarray}
\langle S_{21}\rangle & = & \int d^{4}xC_{21}^{\alpha\beta\mu\nu}G_{\alpha\beta\mu\nu}(x,x)\\
\langle S_{22}\rangle & = & \int d^{4}xC_{22}^{\mu\nu}G_{\mu\nu}(x,x)\\
\langle S_{1}^{2}\rangle & = & \int d^{4}x\int d^{4}x'C_{11}^{\alpha\beta\mu\nu}C_{11}^{\rho\sigma\lambda\tau}G_{\alpha\beta\rho\sigma}(x,x')\partial_{\mu}\partial_{\lambda}^{'}G_{\nu\tau}(x,x')
\end{eqnarray}
where we have neglected the corrections in the $\langle S_{1}^{2}\rangle$
from the term $C_{12}^{\alpha\beta\mu}$ which contributes to high order
operators. $S_{13}$ and $S_{23}$ give vanishing quadratic divergences
as we have explained in the previous section. Thus we can expand the gauge and
graviton propagators into
\begin{eqnarray}
G_{\mu\nu}(p) & = & \frac{\delta_{\mu\nu}}{p^{2}-v\Lambda}+(2\zeta-1)\frac{p_{\mu}p_{\nu}}{(p^{2}-v\Lambda)(p^{2}-2\zeta v\Lambda)}\nonumber \\
 & = & \frac{\delta_{\mu\nu}}{p^{2}}+(2\zeta-1)\frac{p_{\mu}p_{\nu}}{p^{4}}+\mathcal{O}(\Lambda)
\end{eqnarray}
 and
\begin{eqnarray}
G_{\rho\sigma\lambda\tau}(p) & = & \frac{\delta_{\rho(\lambda}\delta_{\tau)\sigma}-\frac{1}{2}\delta_{\rho\sigma}\delta_{\lambda\tau}}{\left(p^{2}-2\Lambda\right)}+(\kappa^{2}\xi-1)\frac{\delta_{\rho(\lambda}p_{\tau)}p_{\sigma}+\delta_{\sigma(\lambda}p_{\tau)}p_{\rho}}{\left(p^{2}-2\Lambda\right)\left(p^{2}-2\kappa^{2}\xi\Lambda\right)}\nonumber \\
 & = & \frac{\delta_{\rho(\lambda}\delta_{\tau)\sigma}-\frac{1}{2}\delta_{\rho\sigma}\delta_{\lambda\tau}}{p^{2}}+(\kappa^{2}\xi-1)\frac{\delta_{\rho(\lambda}p_{\tau)}p_{\sigma}+\delta_{\sigma(\lambda}p_{\tau)}p_{\rho}}{p^{4}}+\mathcal{O}(\Lambda)
\end{eqnarray}
Since the cosmological constant $\Lambda$ is of mass-dimension two,
the corrections arising from $\mathcal{O}(\Lambda)$ are only logarithmically
divergent, we shall not consider its effects below. When the calculation involves
propagators $G_{\mu\nu}(p-q)$ and $G_{\rho\sigma\lambda\tau}(p-q)$
which depend on the external momentum $q$, we will treat respectively $G_{\mu\nu}(p-q)$
as $G_{\mu\nu}(p)$ and $G_{\rho\sigma\lambda\tau}(p-q)$ as $G_{\rho\sigma\lambda\tau}(p)$,
since the $q$-dependent contributions can be regarded as the higher order terms,
like $\partial^{\mu}F_{\alpha\beta}\partial_{\mu}F^{\alpha\beta}$.
With this consideration and approximation, all the remaining leading-contributions will only involve with the following
quadratically divergent tensor- and scalar-type loop integrals
\begin{equation}
\mathcal{I}_{2\mu\nu}=\int d^{4}p\frac{p_{\mu}p_{\nu}}{p^{4}};\quad\mathcal{I}_{2}=\int d^{4}p\frac{1}{p^{2}}.\label{Eq.Tensor+Scalar}
\end{equation}
 In general, one needs a consistent regularization to make the quadratically
divergent integrals well-defined. Without involving the details of
regularization schemes, one can always relate the regularized tensor-type
integral with the regularized scalar-type integral via the general
Lorentz structure as follows
\begin{equation}
\mathcal{I}_{2\mu\nu}^{R}=a_{2}\delta_{\mu\nu}\mathcal{I}_{2}^{R}.\label{Eq.TensorScalar}
\end{equation}
where the superscript $R$ denotes the regularized integral. Here
$a_{2}$ may be different in different regularization schemes. However,
by explicitly calculating one loop diagrams of gauge theories, it
has been shown \cite{YLW} that a consistency condition with
\begin{equation}
a_{2}=\frac{1}{2}\label{eq.consistentc}
\end{equation}
is required to preserve gauge invariance for $\mathcal{I}_{2}^{R}\neq0$.
In the LORE method\cite{YLW}, the condition Eq.~(\ref{eq.consistentc}) is satisfied,
while the naive cut-off regularization does not lead to the condition Eq.~(\ref{eq.consistentc}) as it results to $a_{2}=\frac{1}{4}$. The dimensional regularization is known to suppress the quadratic divergence and gives $\mathcal{I}_{2}^{R}=0$, which leads no quadratic divergence.

With the above general relation, the quadratically divergent parts
of effective action, without including ghost's contribution at one-loop level, are found
to be
\begin{eqnarray}
\langle S_{2}\rangle & = & \kappa^{2}(C_{21}+C_{22})\mathcal{I}_{2}^{R}\frac{1}{4}\int d^{4}x\bar{F}^{2}\nonumber \\
\langle S_{1}^{2}\rangle & = & \kappa^{2}C_{11}\mathcal{I}_{2}^{R}\frac{1}{4}\int d^{4}x\bar{F}^{2}
\end{eqnarray}
where the numerical factors are explicitly given by
\begin{eqnarray}
C_{21} & = & \frac{1}{2}\Bigl([v(1-4a_{2})+6a_{2}](\kappa^{2}\xi-1)+3\Bigl)\nonumber \\
C_{22} & = & \frac{v}{8}(4a_{2}-1)(2\zeta-1)+\frac{\omega^{2}}{\kappa^{2}\xi}\Bigl[(2\zeta-1)a_{2}+1\Bigl]\nonumber \\
C_{11} & = & \frac{2\omega^{2}}{\kappa^{2}\xi}([2\zeta-1]a_{2}+1)+2\kappa^{2}\xi(1-a_{2})+6a_{2}-4\omega(1-a_{2})
\end{eqnarray}
Thus the effective action with the total gravitational field contributions
is found at the one-loop order to be
\begin{eqnarray}
\Gamma_{G} & = & \langle S_{2}\rangle-\frac{1}{2}\langle S_{1}^{2}\rangle=\kappa^{2}C_{G}\mathcal{I}_{2}^{R}\frac{1}{4}\int d^{4}x\bar{F}^{2}\\
C_{G} & = & \frac{(4a_{2}-1)}{8}\Bigl(v\left[(2\zeta-1)-4(\kappa^{2}\xi-1)\right]\nonumber \\
 & + & 8(\kappa^{2}\xi-1)-16\omega-4\Bigl)+6\omega a_{2}
\end{eqnarray}
Thanks the cancelation of $1/\xi$ terms in $\left\langle S_{2}\right\rangle $ and $\left\langle S_{1}^{2}\right\rangle $
, otherwise it would be inconsistent when going back to the Landau-DeWitt gauge $\xi\rightarrow0$. The ghost's contribution to the effective action at the one-loop order can be written as
\begin{equation}
\Gamma_{GH}=\langle S_{GH2}\rangle-\frac{1}{2}\langle S_{GH1}^{2}\rangle
\end{equation}
An additional sign has to be taken care for a ghost loop in the calculation,
the quadratically divergent contributions are found to be
\begin{eqnarray}
\langle S_{GH2}\rangle & = & -\kappa^{2}\omega\mathcal{I}_{2}^{R}\frac{1}{4}\int d^{4}x\bar{F}^{2}\ ;\quad\left\langle S_{GH1}^{2}\right\rangle =0
\end{eqnarray}
which is independent of $a_{2}$ in Eq.~(\ref{Eq.TensorScalar}).

Thus the total quadratically divergent one-loop gravitational contribution
to the effective action has the following gauge invariant form
\begin{eqnarray}
\Gamma & = & \frac{1}{4}\int d^{4}x\bar{F}^{2}+\kappa^{2}C\mathcal{I}_{2}^{R}\frac{1}{4}\int d^{4}x\bar{F}^{2}\label{Eq.EAfinal}
\end{eqnarray}
 where the constant $C$ is given by
\begin{eqnarray}
C & = & C_{G}-\omega=\frac{4a_{2}-1}{8}\Bigl(v\left[(2\zeta-1)-4(\kappa^{2}\xi-1)\right]\nonumber \\
 & + & 8(\kappa^{2}\xi-1)-16\omega-4\Bigl)+\omega(-1+6a_{2})
\end{eqnarray}
Thus the corresponding counter-term is needed to renormalize the gauge field and gauge coupling constant. The renormalized gauge action is given by
\begin{eqnarray}
\Gamma &=& \frac{1}{4}(1 + \delta_2) \int d^4x \bar{F}_{\mu\nu}\bar{F}^{\mu\nu}
\end{eqnarray}
where $\delta_{2}$ is determined via the cancelation of the quadratic divergence $\delta_2 + \kappa^2 C \mathcal{I}^{R}_2 \simeq 0$, namely
\begin{equation}
 \delta_2 \simeq - \kappa^2 C \mathcal{I}^{R}_2
\end{equation}
as the charge renormalization constant $z_e$ is connected to the gauge field renormalization constant $z_2=1+\delta_2 $ via the identity $z_ez^{1/2}_{2}=1$, the gravitational correction to the $\beta$ function is defined to be
\begin{eqnarray}
\beta^{\kappa}_{e}= \mu \frac{\partial}{\partial\mu}e=\mu \frac{\partial}{\partial\mu} z^{-1}_e e^{0}
         =\frac{1}{2}e^0\mu \frac{\partial}{\partial\mu} \delta_2
\end{eqnarray}
from which we can obtain the gravitational corrections to the $\beta$ function
\begin{equation}
\beta_{e}^{\kappa}=\frac{\mu^{2}}{16\pi^{2}}e\kappa^{2}C,\label{Eq.Betafunc}
\end{equation}
where $e$ is the electric charge. Such a result indicates that there do exist quadratically divergent gravitational
contributions to the gauge coupling constant for $C \neq 0$. Let us now impose the Landau-DeWitt gauge condition $v=1,\ \omega=1,\ \zeta=0,\ \xi=0$, and take the gauge invariance consistency condition $a_{2}=1/2$, which leads to a nonzero value for the constant $C=6a_{2}-1-25(a_{2}/2-1/8)=-9/8$ and results in a negative $\beta$ function 
\begin{equation}
\beta_{e}^{\kappa}=-\frac{9\mu^{2}}{128\pi^{2}}e\kappa^{2}
\end{equation}

We would like to emphasize that the above result is gauge condition independent ensured by the Vilkovisky-DeWitt
formalism, and is also independent of any specific regularization
schemes as long as the regularization schemes preserve gauge symmetry and divergent behavior. We then
arrive at the statement that gravity does provide power-law contributions to
the gauge coupling constant and tends to make gauge coupling asymptotically
free.

We are also in the position to make comments on the regularization scheme dependence.
In the dimension regularization, one has $\mathcal{I}_{2}^{R}=0$,
so that $\delta_{2}=0$ and $\beta_{e}^{\kappa}=0$, it is then manifest
that there is no quadratically divergent gravitational contributions
in any case based on the dimensional regularization. In the cut-off
regularization, one has $a_{2}=1/4$ and $C=1/2$, namely $\beta_{e}^{\kappa}=\mu^{2}/(32\pi^{2})e\kappa^{2}$ which leads to no asymptotic freedom.

As a consistent check, let us revisit the traditional background
field method in the harmonic gauge, which is recovered by simply taking
$v=0,\ \omega=0,\ \zeta=1/2,\ \xi=1/\kappa^{2}$ in the above Vilkovisky-DeWitt
formalism. As a consequence, it leads to
\begin{equation}
C=1/2-2a_{2}
\end{equation}
which becomes manifest that in the cut-off regularization, one has
$a_{2}=1/4$, $C=0$ and $\beta_{e}^{\kappa}=0$, which confirms the
previous results given in \cite{Pietrykowski,Ebert,TangWu}. In the
LORE method $a_{2}=1/2$, we have $C=-1/2$ and $\beta_{e}^{\kappa}=-\mu^{2}/(32\pi^{2})e\kappa^{2}$ which confirms our previous result given in Eq.~(\ref{eq:Diagrambeta}) and ref. \cite{TangWu}. Namely, the quadratically divergent gravitational contribution to the
gauge coupling constant is asymptotic free in the traditional background
field or equivalently in the diagrammatic method with the harmonic gauge.

\subsection{Quartic divergences\label{sec:Quartic}}

In this section, we shall restrict ourselves to the quartic divergences
that may appear in the calculation. The quartic divergences have a
form of
\begin{equation}
\bar{A}^{\mu}\bar{A}_{\mu}\int\frac{d^{4}p}{(2\pi)^{4}}1=\bar{A}^{\mu}\bar{A}_{\mu}\mathcal{I}_{4}.\label{eq:QuarticIntegral}
\end{equation}
such a term will violate the $U(1)$ gauge symmetry without adopting a proper regularization scheme to handle it.

In the gravity-gauge sector, it is easy to check that quartic divergences
only show in the contributions from $\left\langle S_{13}^{2}\right\rangle $
and $\left\langle S_{23}\right\rangle $. Both contributions are proportional
to $\omega^{2}/\xi$,
\begin{eqnarray}
\frac{1}{2}\left\langle S_{13}^{2}\right\rangle _{4} & = & \frac{\omega^{2}}{2\kappa^{2}\xi^{2}}\int d^{4}x\int d^{4}x'C_{13}^{\mu\alpha\beta}C_{13}^{\nu\rho\sigma}\partial_{\mu}\partial_{\nu}^{'}G_{\alpha\beta\rho\sigma}(x,x')\partial^{\gamma}\partial^{'\tau}G_{\gamma\tau}(x,x')
\end{eqnarray}
with $C_{13}^{\mu\alpha\beta}=\delta^{\mu(\alpha}\bar{A}^{\beta)}-\frac{1}{2}\delta^{\alpha\beta}\bar{A}^{\mu}$
and $\partial^{\gamma}\partial^{'\tau}G_{\gamma\tau}(x,x')=2\zeta\delta(x,x')$.
Here the subscript $4$ indicates that we are dealing with quartic
divergence. We can show that only the terms proportional to $\kappa^{2}\xi$
in the gravity propagator contribute to the quartic divergent term, 
\begin{eqnarray}
\frac{1}{2}\left\langle S_{13}^{2}\right\rangle _{4} & = & \frac{\omega^{2}}{4\xi}2\zeta\int d^{4}x\bar{A}^{\mu}\bar{A}_{\mu}\mathcal{I}_{4}
\end{eqnarray}
 and similarly we have
\begin{eqnarray}
\left\langle S_{23}\right\rangle _{4} & = & \frac{\omega^{2}}{4\xi}\int d^{4}xC_{23}\partial^{\gamma}\partial^{\tau}G_{\gamma\tau}(x,x)\nonumber \\
 & = & \frac{\omega^{2}}{4\xi}2\zeta\int d^{4}x\bar{A}^{\mu}\bar{A}_{\mu}\mathcal{I}_{4}
\end{eqnarray}
Luckily, the cancelation occurs that $\left\langle S_{23}\right\rangle _{4}-\frac{1}{2}\left\langle S_{13}^{2}\right\rangle _{4}=0$.
If they could not cancel each other, we would get inconsistent result in the limit of $\xi\rightarrow0$. Again, we emphasize that we only
confine our discussion here to quartic divergence. In a general gauge
for arbitrary $\omega$, the connection term $S_{T}$ in Eq.(\ref{eq:S_T})
should be included as well. Since $S_{T}$ involves
$S_{,i}$, Eq. (\ref{eq:FirstDflat}) has only $\bar{F}_{\mu\nu}$
or $\partial_{\mu}\bar{F}^{\mu\nu}$ in the absence of cosmological constant
$\Lambda$, there will be no quartically divergent correction to $\bar{A}_{\mu}\bar{A}^{\mu}$
from $S_{T}$.

However, in the ghost sector there is no such a cancelation. The second term
$\omega\bar{c}^{\lambda}\bar{A}_{\lambda}\square c$ in eq.(\ref{eq:GhostSGH1}),
and the terms in the second bracket of eq.(\ref{eq:GhostSGH2}), $-\omega\left[\bar{c}^{\lambda}\bar{A}_{\lambda}\bar{A}_{\rho}\square c^{\rho}+\bar{c}^{\lambda}\bar{A}_{\lambda}\bar{A}_{\nu,\rho}c^{\rho,\nu}\right]$,
originate from the interaction term $\omega\bar{A}_{\lambda}\partial^{\mu}a_{\mu}$ in
the Landau-DeWitt gauge condition for graviton in eq.(\ref{eq:GaugeForGraviton}).
These two terms will give a non-zero quartic divergence.
\begin{eqnarray}
\frac{1}{2}\left\langle S_{GH1}^{2}\right\rangle _{4} & = & \frac{1}{2}\omega\left\langle \int d^{4}x\int d^{4}x'\bar{c}^{\lambda}\bar{A}_{\lambda}\square c\bar{c}\bar{A}_{\nu}\square c^{\nu}\right\rangle \nonumber \\
 & = & \frac{\kappa^{2}}{4}\omega\int d^{4}x\bar{A}^{\mu}\bar{A}_{\mu}\mathcal{I}_{4}
\end{eqnarray}
 and
\begin{eqnarray}
\left\langle S_{GH2}\right\rangle _{4} & = & -\omega\left\langle \int d^{4}x\bar{c}^{\lambda}\bar{A}_{\lambda}\bar{A}_{\nu}\square c^{\nu}\right\rangle \nonumber \\
 & = & -\frac{\kappa^{2}}{2}\omega\int d^{4}x\bar{A}^{\mu}\bar{A}_{\mu}\mathcal{I}_{4}
\end{eqnarray}
where a sign has been added for a ghost loop. It is seen that the total contribution to the effective action is nonzero $\left\langle S_{GH2}\right\rangle _{4}-\frac{1}{2}\left\langle S_{GH1}^{2}\right\rangle _{4}\neq0$, which leads to a divergent mass term and violates $U(1)$ gauge invariance without imposing proper regularization schemes to treat such a quartic divergence. Obviously, in the cut-off regularization, $\mathcal{I}_{4}^{R}$ is proportional to $\Lambda^{4}$, which then destroys the gauge invariance. While the LORE method is found to be a proper regularization scheme as it leads $\mathcal{I}_{4}^{R} = 0$, so that the regularized quartic divergence disappears and the gauge invariance is maintained. Though the dimensional regularization results in $\mathcal{I}_{4}^{R} = 0$, while it also gives $\mathcal{I}_{2}^{R} = 0$.

\section{Conclusions\label{sec:Summary}}

In summary, we have investigated the one-loop quadratically divergent gravitational
corrections to gauge Green's functions both in diagrammatic calculation and in the gauge condition
independent Vilkovisky-DeWitt background field method. As a consequence, we have obtained in both cases the quadratically divergent gravitational contributions to the $\beta$ function of gauge coupling constant. We limit our discussion in one-loop approximation. This approximation can break down as approaching the Planck scale, where new framework of quantum gravity is needed. 

In the diagrammatic approach, we have explicitly performed the calculations for the two-, three- and four-point gauge Green's functions at one-loop level with graviton contributions. We have demonstrated for the first time that the Slavnov-Taylor identities are satisfied for these gravitational corrections, which is found to be irrespective of the regularization schemes. However, our analysis has shown that the gravitational contribution to the $\beta$ function is dependent on the regularization schemes in the harmonic gauge condition. Both the cut-off and dimensional regularization schemes lead to a zero result, here the former is due to the accidental cancelation with the inconsistency relation $\mathcal{I}^R_{2\mu\nu} = \frac{1}{4}g_{\mu\nu} \mathcal{I}^R_{2}$ which spoils gauge invariance, and the latter is due to the well-known suppression effect of dimensional regularization to the quadratic divergence with $ \mathcal{I}^R_{2}=0$. In contrast, the LORE method gives a non-zero result that render all gauge theory asymptotically free at very high energy scale, this is because the LORE method preserves the gauge symmetry and maintains the quadratic divergent behavior with the consistency condition $\mathcal{I}^R_{2\mu\nu} = \frac{1}{4}g_{\mu\nu} \mathcal{I}^R_{2}$.

In the second part of our calculations, we have worked within the framework of Vilkovisky-DeWitt's effective action which is gauge-condition independent. We have shown in a regularization scheme independent way that there is in general quadratically divergent gravitational contributions to the gauge coupling. It is interesting to notice that when reducing to the framework of traditional background field approach with harmonic gauge condition, we arrive at the same results obtained in the diagrammatic approach, which shows the equivalence between the diagrammatic approach and traditional background method. We have found that in any case the symmetry-maintaining and divergent-behavior-preserving LORE method leads to an asymptotic free power-law running of gauge coupling at one-loop near the Planck scale due to quantum gravitational contributions. In particular, we have paid attention to the treatment on the quartic divergent effect which in general violates gauge invariance, again the LORE method is found to be a proper regularization scheme to handle the quartic divergence for ensuring gauge invariance.

{\it Note added}: At the final stage of this work, there is a preprint \cite{Nielsen}, which also discussed the Einstein-Maxwell system, Ward identities and Vilkovisky-DeWitt's formalism. Part of our calculation on quadratic divergence is confirmed by \cite{Nielsen} using proper-time representation.

\centerline{{\bf Acknowledgement}}

\vspace{20 pt}

This work was supported in part by the
National Science Foundation of China (NSFC) under Grant \#No.
10821504, 10975170 and the key project of the Chinese Academy of Sciences.

\newpage

\appendix{\bf Appendix}

In the following appendix sections, we are going to present a concise overview for the traditional and Vilkovisky-DeWitt's
modified effective action for completeness and convenience. The introduction of traditional effective
action can be found in the standard textbook\cite{WeinbergBook}. The
original idea of Vilkovisky-DeWitt's effective action was presented
in\cite{Vilkovisky,DeWitt}. For a pedagogical review on this topic,
the readers are referred to the article\cite{Tomsbook} and references therein.

\section{Traditional effective action}

We begin with a brief description of the traditional approach to define
an effective action for non-gauge theories. Let $S[\varphi]$ be the
classical action for the theory, then the generating functional $Z[J]$
in the presence of external currents for the n-point Green's functions
is defined by
\begin{equation}
Z[J]=\mathcal{N}\int\mathcal{D}\varphi\exp{i\left\{ S[\varphi]+\varphi^{i}J_{i}\right\} }\label{eq:Zfunctional}
\end{equation}
 where the functional integration measure is
\begin{equation}
\mathcal{D}\varphi=\left(\prod_{i}d\varphi^{i}\right)
\end{equation}
 $\mathcal{N}$ is an irrelevant constant for normalization and will
be neglected below. $Z[J]$ has a physical diagrammatic picture that
it is the the sum of all vacuum-to-vacuum amplitudes, including both
disconnected and connected diagrams. For the connected Green's functions,
it is useful to define another generating functional, $W[J]$ , with
\begin{equation}
W[J]=-i\ln{Z[J]}
\end{equation}
 For One-Particle-Irreducible(OPI) diagrams, one can go one step further.
Define the background fields $\bar{\varphi}^{i}$ as
\begin{eqnarray}
\bar{\varphi}^{i}(x) & \equiv & \frac{\left\langle \text{out}\left|\varphi^{i}(x)\right|\text{in}\right\rangle _{J}}{\left\langle \text{out}|\text{in}\right\rangle _{J}}=\frac{-i}{Z[J]}\frac{\delta Z[J]}{\delta J^{i}(x)}=\frac{\delta W[J]}{\delta J^{i}(x)}
\end{eqnarray}
 Now, the quantum effective action $\Gamma[\bar{\varphi}]$ is defined
by the Legendre transformation of $W[J]$,
\begin{equation}
\Gamma[\bar{\varphi}]\equiv W[J]-\bar{\varphi}^{i}J_{i}
\end{equation}
 It can be shown that $\bar{\varphi}^{i}$ satisfies the equation
\begin{equation}
\frac{\delta\Gamma[\bar{\varphi}]}{\delta\bar{\varphi}^{i}(x)}=-J_{i}(x)
\end{equation}
 and $\Gamma[\bar{\varphi}]$ is the generating functional for OPI
Green functions. In the functional integral representation, we have
\begin{eqnarray}
\exp{i\Gamma[\bar{\varphi}]} & = & \int\mathcal{D}\varphi\exp{i\left\{ S[\varphi]+\left(\varphi^{i}-\bar{\varphi}^{i}\right)J_{i}\right\} }\nonumber \\
 & = & \int\mathcal{D}\varphi\exp{i\left\{ S[\varphi]-\left(\varphi^{i}-\bar{\varphi}^{i}\right)\frac{\delta\Gamma[\bar{\varphi}]}{\delta\bar{\varphi}^{i}}\right\} }
\end{eqnarray}
Both sides of the above equation have $\Gamma[\bar{\varphi}]$, so
it will involve an iterative procedure to solve the equation in perturbative
expansion. For example, to get $\Gamma[\bar{\varphi}]$ at one loop
level on the Left-Hand-Side(LHS), we can replace the $\Gamma[\bar{\varphi}]$
on the Right-Hand-Side(RHS) with its tree level value $S[\bar{\varphi}]$.

In the background field approach, one expands the fields $\varphi^{i}$
as the sum of background fields $\bar{\varphi}^{i}$ and quantum fields
$\eta^{i}$,
\begin{equation}
\varphi^{i}=\bar{\varphi}^{i}+\eta^{i}.
\end{equation}
 To get the one-loop effective action, only the quadratic terms of
$\eta^{i}$ in the exponent need to be kept,
\begin{eqnarray*}
\exp{i\Gamma[\bar{\varphi}]} & = & \int\mathcal{D}\eta\exp{i\left\{ S[\bar{\varphi}^{i}+\eta^{i}]-\eta^{i}\frac{\delta S[\bar{\varphi}]}{\delta\bar{\varphi}^{i}}\right\} }\\
 & = & \exp{iS[\bar{\varphi}]}\int\mathcal{D}\eta\exp{i\left\{ \frac{1}{2}\eta^{i}\frac{\delta S[\bar{\varphi}]}{\delta\bar{\varphi}^{i}\delta\bar{\varphi}^{j}}\eta^{j}\right\} }
\end{eqnarray*}
 then, the effective action $\Gamma[\bar{\varphi}]$ is given by
\begin{eqnarray}
\Gamma[\bar{\varphi}] & = & S[\bar{\varphi}]+\frac{i}{2}\ln\det{S_{,ij}}\label{eq:effaction1}
\end{eqnarray}

\section{The Vilkovisky-DeWitt effective action}

The above formalism for defining quantum effective action has the
problem that it depends on the parametrization of $\varphi'=\varphi'(\varphi)$
\cite{Vilkovisky}. Suppose $S[\varphi]$ is a scalar under the transformation
$\varphi'=\varphi'(\varphi)$, so should be expected for $\Gamma[\varphi]$.
However, in a different parametrization $\varphi'=\varphi'(\varphi)$,
the effective action will be changed \cite{Vilkovisky}
\begin{equation}
\Gamma'[\bar{\varphi'}]=S'[\bar{\varphi}']+\frac{i}{2}\ln\det{\left[S_{,ij}+S_{,k}\frac{\partial^{2}\varphi^{k}}{\partial\varphi'^{l}\partial\varphi'^{m}}\frac{\partial\varphi'^{l}}{\partial\varphi^{i}}\frac{\partial\varphi'^{m}}{\partial\varphi^{j}}\right]}\label{eq:effaction2}
\end{equation}
Since $S'[\bar{\varphi}']=S[\bar{\varphi}]$, eq.(\ref{eq:effaction2})
will be different from eq.(\ref{eq:effaction1}) at one loop level already.
To solve the field parametrization dependence, Vilkovisky \cite{Vilkovisky}
suggested that we might regard the field space $\varphi^{i}$ as a
manifold $M$ which is associated with the metric $g_{ij}[\varphi]$, connection
$\Gamma_{jk}^{i}$, treat the field $\varphi^{i}$ as the coordinates on this
manifold,  and define the effective action in terms of parametrization
invariant quantities. Later, DeWitt discussed this issue further in
\cite{DeWitt}. In the following, we shall follow the discussion in ref.
\cite{Tomsbook} where a complete and clear discussion was given.

Define the two-point function or world function \cite{DeWittbook},
\begin{equation}
\sigma[\varphi_{\star};\varphi]=\frac{1}{2}(\text{length of geodesic from \ensuremath{\varphi_{\star}}to \ensuremath{\varphi}})^{2}
\end{equation}
 and
\begin{equation}
\sigma^{i}[\varphi_{\star};\varphi]=g^{ij}[\varphi_{\star}]\frac{\delta\sigma[\varphi_{\star};\varphi]}{\delta\varphi_{\star}^{j}}
\end{equation}
 $\sigma^{i}[\varphi_{\star};\varphi]$ is a vector and tangent to
the geodesic line on the field space that connects $\varphi_{\star}$
and $\varphi$. $\sigma^{i}[\varphi_{\star};\varphi]$ can be expanded
in powers of $\eta^{i}=\varphi^i-\bar{\varphi}^{i}$,
\begin{equation}
\sigma^{i}[\varphi_{\star};\varphi]=-\eta^{i}+\sum_{n=2}^{\infty}\frac{1}{n!}\sigma^{i}{}_{j_{1}\cdots j_{n}}\eta^{j_{1}}\cdots\eta^{j_{n}},\:\sigma^{i}{}_{jk}=\Gamma_{jk}^{i}
\end{equation}
 where $\varphi_{\star}$ is an arbitrary point at the moment. Then
the generating functional eq.(\ref{eq:Zfunctional}) is modified to be
\begin{eqnarray}
Z[J] & = & \int d\mu[\varphi_{\star};\varphi]\exp{i\left\{ S[\varphi]-J_{i}\sigma^{i}[\varphi_{\star};\varphi]\right\} }=\exp{iW[J;\varphi_{\star}]}\label{eq:VDZfunctional}
\end{eqnarray}
 where the measure
\begin{eqnarray*}
d\mu[\varphi_{\star};\varphi] & = & \left(\prod_{i}d\sigma^{i}[\varphi_{\star};\varphi]\right)\sqrt{\left|g[\varphi_{\star}]\right|}=\left(\prod_{i}d\varphi^{i}\right)\sqrt{\left|g[\varphi]\right|}\left|J[\varphi_{\star};\varphi]\right|
\end{eqnarray*}
 $J[\varphi_{\star};\varphi]$ is the Jacobean factor, which is irrelevant
at one loop order. $S[\varphi]$ is treated as functional of $\varphi_{\star}$
and $\sigma^{i}[\varphi_{\star};\varphi]$, $\hat{S}\left[\varphi_{\star};\sigma^{i}[\varphi_{\star};\varphi]\right]$
is defined by covariant Taylor expansion,
\begin{equation}
S[\varphi]=\hat{S}\left[\varphi_{\star};\sigma^{i}[\varphi_{\star};\varphi]\right]=\sum_{n=0}^{\infty}\frac{(-1)^{n}}{n!}S_{;i_{1}\cdots i_{n}}[\varphi_{\star}]\sigma^{i_{1}}[\varphi_{\star};\varphi]\cdots\sigma^{i_{n}}[\varphi_{\star};\varphi]
\end{equation}
then $Z[J]$ of eq.(\ref{eq:VDZfunctional}) can be regarded as the generating
functional for Green function of $\sigma^{i}[\varphi_{\star};\varphi]$,
but we are more interested in Green function of $\varphi^{i}$. Define
$\bar{\varphi}$ and $v^{i}$ through
\begin{equation}
v^{i}\equiv\sigma^{i}[\varphi_{\star};\bar{\varphi}]\equiv\left\langle \sigma^{i}[\varphi_{\star};\varphi]\right\rangle =\frac{\delta W[J;\varphi_{\star}]}{\delta J_{i}}.
\end{equation}
 Similarly, the corresponding effective action then is
\begin{equation}
\hat{\Gamma}\left[\varphi_{\star};\sigma^{i}[\varphi_{\star};\bar{\varphi}]\right]=W[J;\varphi_{\star}]+J_{i}\sigma^{i}[\varphi_{\star};\bar{\varphi}],
\end{equation}

\begin{equation}
\exp{i\hat{\Gamma}\left[\varphi_{\star};\sigma^{i}[\varphi_{\star};\bar{\varphi}]\right]}=\int d\mu[\varphi_{\star};\varphi]\exp{i\left\{ S[\varphi]-\frac{\delta\hat{\Gamma}}{\delta v^{i}}\left(\sigma^{i}[\varphi_{\star};\varphi]-v^{i}\right)\right\} }.
\end{equation}
 By expanding the functional $\hat{S}\left[\varphi_{\star};\sigma^{i}[\varphi_{\star};\varphi]\right]$
at $\sigma^{i}=v^{i}$
\begin{equation}
\hat{S}\left[\varphi_{\star};\sigma^{i}[\varphi_{\star};\varphi]\right]=\hat{S}\left[\varphi_{\star};v^{i}\right]+\sum_{n=0}^{\infty}\frac{1}{n!}\frac{\delta^{n}\hat{S}\left[\varphi_{\star};v^{i}\right]}{\delta v^{i_{1}}\cdots\delta v^{i_{n}}}(\sigma^{i_{1}}-v^{i_{1}})\cdots(\sigma^{i_{n}}-v^{i_{n}}),
\end{equation}
 where the expansion coefficients are connected to those of covariant
Taylor expansion,
\begin{equation}
\frac{\delta^{n}\hat{S}\left[\varphi_{\star};v^{i}\right]}{\delta v^{i_{1}}\cdots\delta v^{i_{n}}}=(-1)^{n}S_{;i_{1}\cdots i_{n}}[\varphi_{\star}]+\sum_{m=n+1}^{\infty}\frac{(-1)^{m}}{(m-n)!}S_{;i_{1}\cdots i_{m}}[\varphi_{\star}]v^{i_{1}}\cdots v^{i_{m}}
\end{equation}
 Then at one loop level, the effective action is given by
\begin{eqnarray}
\hat{\Gamma}\left[\varphi_{\star};\sigma^{i}[\varphi_{\star};\bar{\varphi}]\right] & = & \hat{S}\left[\varphi_{\star};\sigma^{i}[\varphi_{\star};\bar{\varphi}]\right]+\frac{i}{2}\ln\det{\left[g^{ik}[\varphi_{\star}]\frac{\delta^{2}\hat{S}\left[\varphi_{\star};v^{i}\right]}{\delta v^{k}\delta v^{j}}\right]},
\end{eqnarray}
 When $\varphi^{\star}$ is arbitrary, we can take the limit $\varphi_{\star}\rightarrow\bar{\varphi}$
and $v^{i}\rightarrow0$, then we have
\begin{equation}
\left.\frac{\delta^{n}\hat{S}\left[\varphi_{\star};v^{i}\right]}{\delta v^{i_{1}}\cdots\delta v^{i_{n}}}\right|_{v=0}=(-1)^{n}S_{;i_{1}\cdots i_{n}},
\end{equation}
 and the effective action
\begin{eqnarray}
\Gamma[\bar{\varphi}]=\hat{\Gamma}\left[\bar{\varphi};\sigma^{i}[\bar{\varphi};\bar{\varphi}]\right] & = & \hat{S}\left[\bar{\varphi};\sigma^{i}[\bar{\varphi};\bar{\varphi}]\right]+\frac{i}{2}\ln\det{\left[g^{ik}[\varphi_{\star}]\frac{\delta^{2}\hat{S}\left[\varphi_{\star};v^{i}\right]}{\delta v^{k}\delta v^{j}}\right]_{v=0}}\nonumber \\
 & = & S[\bar{\varphi}]+\frac{i}{2}\ln\det{\left[\nabla^{i}\nabla_{j}S[\bar{\varphi}]\right]}
\end{eqnarray}
 The above formalism can be generalized to multi-loops \cite{Tomsbook}.

For gauge theories, some modifications are needed. Let $S[\varphi]$
represent the classical action functional for a gauge theory, it is
gauge invariant under the transformation
\begin{equation}
\delta\varphi^{i}=K_{\alpha}^{i}[\varphi]\delta\epsilon^{\alpha}.
\end{equation}
with $K_{\alpha}^{i}[\varphi]$ regarded as the generators of gauge
transformations and $\delta\epsilon^{\alpha}$ infinitesimal parameters.
$S[\varphi]$ is gauge invariant in the sense of
\begin{equation}
S_{,i}\delta\varphi^{i}=S_{,i}K_{\alpha}^{i}[\varphi]\delta\epsilon^{\alpha}=0,\forall\quad \delta\epsilon^{\alpha}\Longrightarrow S_{,i}K_{\alpha}^{i}[\varphi]=0
\end{equation}
To quantize gauge theory, a gauge fixing condition has to be imposed,
for instance, $\chi^{\alpha}[\varphi]=f^{\alpha}$, where $f^{\alpha}$
is independent of $\varphi^{i}$. Since we want to fix the gauge field,
the gauge condition then should not be gauge invariant. Then require
$\chi^{\alpha}[\varphi+\delta\varphi]=\chi^{\alpha}[\varphi]$ hold
only if $\delta\epsilon^{\alpha}=0$, one has
\begin{equation}
\chi^{\alpha}{}_{,i}[\varphi]K_{\beta}^{i}[\varphi]\delta\epsilon^{\beta}\equiv Q^{\alpha}{}_{\beta}[\varphi]\delta\epsilon^{\beta}=0
\end{equation}
 Then $\det Q^{\alpha}{}_{\beta}$ is usually called as the Faddeev-Popov
factor \cite{FaddeevPopov}. In Path-Integral quantization, for a
field space $M$ that has gauge symmetry $G$, we only have to integrate
the gauge nonequivalent field configuration on the reduced field space
$\mathcal{M}=M/G$ for the generating functional, and use $\tilde{g}_{ij}$
and $\tilde{\Gamma}_{jk}^{i}$ on $\mathcal{M}$ to define the gauge
and parametrization invariant effective action. However, it is usually
more convenient to work in the whole field space $M$ by inserting
Faddeev-Popov factor $\det Q^{\alpha}{}_{\beta}$ and gauge condition
$\delta\left[\chi^{\alpha}-f^{\alpha}\right]$ into the integral measure
and expressing $\tilde{g}_{ij}$ and $\tilde{\Gamma}_{jk}^{i}$ on
$\mathcal{M}$ in terms of $g_{ij}$ and $\Gamma_{jk}^{i}$ on $M$,
with \cite{Vilkovisky,DeWitt}
\begin{eqnarray}
\tilde{g}_{ij} & = & g_{ij}-K_{i\alpha}\gamma^{\alpha\beta}K_{j\beta},\;\gamma_{\alpha\beta}=K_{\alpha}^{i}g_{ij}K_{\beta}^{j}\\
\tilde{\Gamma}_{ij}^{k} & = & \Gamma_{ij}^{k}+T_{ij}^{k}+K_{\alpha}^{k}A_{ij}^{\alpha}\label{eq:fullconn}\\
T_{ij}^{k} & = & \frac{1}{2}\gamma^{\alpha\lambda}\gamma^{\beta\tau}K_{\alpha i}K_{\beta j}(K_{\lambda}^{n}K_{\tau;n}^{k}+K_{\tau}^{n}K_{\lambda;n}^{k})\nonumber \\
 &  & -\gamma^{\alpha\beta}(K_{\alpha i}K_{\beta;j}^{k}+K_{\alpha j}K_{\beta;i}^{k})\nonumber
\end{eqnarray}
where $A_{ij}^{\alpha}=A_{ji}^{\alpha}$ is arbitrary. $\tilde{\Gamma}_{ij}^{k}$
is different from $\Gamma_{ij}^{k}$ by two additional terms, $T_{ij}^{k}$
and $K_{\alpha}^{k}A_{ij}^{\alpha}$. It will be shown later that
$K_{\alpha}^{k}A_{ij}^{\alpha}$ term will not contribute to one loop
effective action because of the gauge invarance of $S$ and then to
any order $\Gamma$ by induction. Choose a gauge fixing condition,
$\chi^{\alpha}=f^{\alpha}$, then the effective action is
\begin{equation}
\exp{i\hat{\Gamma}\left[\varphi_{\star};v^{i}\right]}=\int d\mu[\varphi_{\star};\varphi]\delta\left[\chi^{\alpha}-f^{\alpha}\right]\det Q^{\alpha}{}_{\beta}\exp{i\left[S[\varphi]-\frac{\delta\hat{\Gamma}}{\delta v^{i}}\left(\sigma^{i}[\varphi_{\star};\varphi]-v^{i}\right)\right]}
\end{equation}
Physical results should be independent of this choice of $f^{\alpha}$,
so we can insert the integration with $\int\mathcal{D}f^{\alpha}\exp\left[\frac{i}{2\Omega}f^{\alpha}f_{\alpha}\right]$
and do a Gaussian average over $f^{\alpha}$. Integrate $f^{\alpha}$
first and use $\delta\left[\chi^{\alpha}-f^{\alpha}\right]$, $\exp\left[\frac{i}{2\Omega}f^{\alpha}f_{\alpha}\right]$
is turned into $\exp\left[\frac{i}{2\Omega}\chi^{\alpha}\chi_{\alpha}\right]$,
then at one-loop approximation the effective action is given by
\begin{equation}
\Gamma[\bar{\varphi}]=S[\bar{\varphi}]-i\ln\det Q^{\alpha}{}_{\beta}+\frac{i}{2}\ln\det\left(\nabla^{i}\nabla_{j}S[\bar{\varphi}]+\frac{1}{2\Omega}\chi_{\alpha}{}^{,i}\chi^{\alpha}{}_{,j}\right)
\end{equation}
 where $\nabla_{i}\nabla_{j}S[\bar{\varphi}]=S_{,ij}[\bar{\varphi}]-\tilde{\Gamma}_{ij}^{k}S_{,k}[\bar{\varphi}]$.
The corresponding effective action with Euclidean metric is
\begin{equation}
\Gamma[\bar{\varphi}]=S[\bar{\varphi}]-\ln\det Q^{\alpha}{}_{\beta}+\frac{1}{2}\ln\det\left(\nabla^{i}\nabla_{j}S[\bar{\varphi}]+\frac{1}{2\Omega}\chi_{\alpha}{}^{,i}\chi^{\alpha}{}_{,j}\right)\label{eq:fullVDaction}
\end{equation}

To calculate the above effective action, one can either use the standard
procedure, $\ln\det D\rightarrow\text{Tr \ensuremath{\ln}}D$ and
expand $\text{\ensuremath{\ln}}D$ in series, or equivalently rewrite
the determinant back to the functional integration \cite{DJToms},
\begin{eqnarray}
\Gamma_{G} & = & \frac{1}{2}\ln\det\left(\nabla^{i}\nabla_{j}S[\bar{\varphi}]+\frac{1}{2\Omega}\chi_{\alpha}{}^{,i}\chi^{\alpha}{}_{,j}\right)=-\ln\int\mathcal{D}\eta e^{-S_{q}}\label{eq:VDaction}\\
S_{q} & = & \frac{1}{2}\eta^{i}\eta^{j}\left[S_{,ij}-\tilde{\Gamma}_{ij}^{k}S_{,k}+\frac{1}{2\Omega}\chi_{\alpha}{}^{,i}\chi^{\alpha}{}_{,j}\right]\\
\Gamma_{GH} & = & -\ln\det Q_{\alpha\beta}=-\ln\int\left[\mathcal{D}\bar{c}\mathcal{D}c\right]e^{-S_{GH}},\label{eq:VDghost}
\end{eqnarray}
with $S_{q}=S_{0}+S_{1}+S_{2}$ in eq.(\ref{eq:VDaction}) and $S_{GH}=\bar{\eta}_{\alpha}Q^{\alpha}{}_{\beta}\eta^{\beta}=S_{GH0}+S_{GH1}+S_{GH2}$
in eq.(\ref{eq:VDghost}). The subscripts on $S$ denotes the order in
the background field $\bar{\varphi}$. $\Gamma_{GH}$ is the ghost
contribution with $\bar{c}_{\alpha}$ and $c^{\beta}$ are anti-commuting
ghost fields. At one loop order, we have
\begin{eqnarray}
\Gamma_{G} & = & -\ln\int\mathcal{D}\eta e^{-S_{0}-S_{1}-S_{2}}=-\ln\int\mathcal{D}\eta\left[1-S_{2}+\frac{1}{2}S_{1}^{2}\right]e^{-S_{0}}\nonumber \\
 & = & -\ln\int\mathcal{D}\eta e^{-S_{0}}+\frac{\int\mathcal{D}\eta\left[S_{2}-\frac{1}{2}S_{1}^{2}\right]e^{-S_{0}}}{\int\mathcal{D}\eta e^{-S_{0}}}\nonumber \\
 & \thickapprox & \left\langle S_{2}\right\rangle -\frac{1}{2}\left\langle S_{1}^{2}\right\rangle
\end{eqnarray}
 where $\thickapprox$ means that we have ignored the irrelevant infinite
constant. Similarly, for the ghost's part, we have
\begin{eqnarray}
\Gamma_{GH} & = & -\ln\int\left[\mathcal{D}\bar{c}\mathcal{D}c\right]e^{-S_{GH0}-S_{GH1}-S_{GH2}}\nonumber \\
 & \thickapprox & \left\langle S_{GH2}\right\rangle -\frac{1}{2}\left\langle S_{GH1}^{2}\right\rangle
\end{eqnarray}
 Note that the connection terms $T_{ij}^{k}S_{,k}$ and $A_{ij}K_{\alpha}^{k}S_{,k}$
in eq.(\ref{eq:VDaction}) can be written as \cite{Huggins}
\begin{eqnarray}
S_{T} & = & -\frac{1}{2}\eta^{i}T_{ij}^{k}S_{,k}\eta^{j}=(\eta^{i}K_{i}^{\beta})K_{\beta;j}^{k}S_{,k}(\eta^{j}-\frac{1}{2}\eta^{l}K_{l}^{\alpha}K_{\alpha}^{j})\label{eq:S_T}\\
S_{K} & = & -\frac{1}{2}\eta^{i}A_{ij}K_{\alpha}^{k}S_{,k}\eta^{j}=0,\:\text{since }K_{\alpha}^{k}S_{,k}=0\label{eq:S_K}
\end{eqnarray}
For the sake of $S_{T}$, we will work in Landau-DeWitt gauge condition
\cite{FradkinTseytlin} which has the following feature and can simplify
the calculation significantly
\begin{equation}
\chi_{\alpha}=K_{\alpha i}[\bar{\varphi}]\eta^{i}=0\Longrightarrow S_{T}=0\label{eq:LDgauge}
\end{equation}
This means that the difference between $\tilde{\Gamma}_{ij}^{k}$
and $\Gamma_{ij}^{k}$ does not contribute to the effective action
at one loop. For multi-loop result, this is only true for special
case that the metric $g_{ij}$ doesn't depend on the field $\varphi^{i}$
\cite{FradkinTseytlin,Tomsbook}. In this gauge, we can use the representation
of $\delta$-function,
\begin{equation}
\delta[\chi^{\alpha}]=\lim_{\Omega\rightarrow0}\left[\det\left(\frac{\delta_{\alpha\beta}}{4\pi\Omega}\right)\right]^{\frac{1}{2}}\exp\left[-\frac{1}{2\Omega}\chi^{\alpha}\chi_{\alpha}\right]
\end{equation}
 Then at one-loop order with Landau-DeWitt gauge, the effective action
is given by
\begin{eqnarray}
\Gamma[\bar{\varphi}] & = & S[\bar{\varphi}]-\ln\det Q_{\alpha\beta}[\bar{\varphi}]\nonumber \\
 &  & +\frac{1}{2}\lim_{\Omega\rightarrow0}\ln\det\left(\nabla^{i}\nabla_{j}S[\bar{\varphi}]+\frac{1}{2\Omega}K_{\alpha}^{i}[\bar{\varphi}]K_{j}^{\alpha}[\bar{\varphi}]\right)\label{Eq:VDaction}
\end{eqnarray}
 with $\nabla_{i}\nabla_{j}S[\bar{\varphi}]=S_{,ij}[\bar{\varphi}]-\Gamma_{ij}^{k}S_{,k}[\bar{\varphi}]$,
here the Christoffel connection $\Gamma_{ij}^{k}$ is determined by
$g_{ij}[\varphi]$. Note that if any other gauge condition is chosen,
Eq.~({\ref{Eq:VDaction}}) will not be true and the complicated
form will replace it with the full $\tilde{\Gamma}_{ij}^{k}$, Eq.($\ref{eq:fullconn}$).
It is noticed that the connection term $\Gamma_{ij}^{k}S_{,k}[\bar{\varphi}]$
distinguishes the Vilkovisky-DeWitt's method from the traditional
background-field method. Also, $\Omega$ in Landau-DeWitt gauge has
to be enforced to $0$ at the end of calculation since it has a different
origin from the $\Omega$ in Eq.($\ref{eq:fullVDaction}$).

In this appendix, we shall give the useful formula in our calculation.
We also show the details of our computation of the Christoffel connection
in the field space, $\Gamma_{ij}^{i}$ and the functional derivatives,
$S_{,i}$ and $S_{,ij}$ in a general background space-time. The classical
action functional of Einstein-Maxwell theory with Euclidean metric
is
\begin{equation}
S=S_{M}+S_{G}=\int d^{4}x|g(x)|^{\frac{1}{2}}\left[\frac{1}{4}F_{\mu\nu}F^{\mu\nu}-\frac{2}{\kappa^{2}}(R-2\Lambda)\right],\label{eq:GravityL}
\end{equation}
 with $F_{\mu\nu}=\partial_{\mu}A_{\nu}-\partial_{\nu}A_{\mu}$ and
$\kappa^{2}=32\pi G$, $G$ is the Newton's gravitational constant,
$\Lambda$ is the cosmological constant, and
\begin{eqnarray}
S_{M} & = & \frac{1}{4}\int d^{4}x|g(x)|^{\frac{1}{2}}F_{\mu\nu}F^{\mu\nu},\\
S_{G} & = & -\frac{2}{\kappa^{2}}\int d^{4}x|g(x)|^{\frac{1}{2}}(R-2\Lambda),
\end{eqnarray}
 and Riemann tensor
\begin{eqnarray}
R^{\rho}{}_{\sigma\mu\nu} & = & \partial_{\mu}\Gamma_{\nu\sigma}^{\rho}-\partial_{\nu}\Gamma_{\mu\sigma}^{\rho}+\Gamma_{\mu\lambda}^{\rho}\Gamma_{\nu\sigma}^{\lambda}-\Gamma_{\mu\lambda}^{\rho}\Gamma_{\nu\sigma}^{\lambda},\label{eq:RiemannTensor}\\
\Gamma_{\mu\nu}^{\rho} & = & \frac{1}{2}g^{\rho\lambda}\left[\frac{\partial g_{\mu\lambda}}{\partial x^{\nu}}+\frac{\partial g_{\nu\lambda}}{\partial x^{\mu}}-\frac{\partial g_{\mu\nu}}{\partial x^{\lambda}}\right].
\end{eqnarray}
Although we use the same symbol $\Gamma_{\mu\nu}^{\rho}$, it should
not be confused with the connection $\Gamma_{jk}^{i}[\varphi^{i}]$
on the field space. The action eq.(\ref{eq:GravityL}) is invariant under
general coordinate and $U(1)$ gauge transformations,
\begin{eqnarray}
\delta g_{\mu\nu} & = & -\delta\epsilon^{\rho}g_{\mu\nu,\rho}-\delta\epsilon^{\rho}{}_{,\mu}g_{\rho\nu}-\delta\epsilon^{\rho}{}_{,\nu}g_{\rho\mu},\\
\delta A_{\mu} & = & -\delta\epsilon^{\nu}A_{\mu,\nu}-\delta\epsilon^{\nu}{}_{,\mu}A_{\nu}+\delta\epsilon_{,\mu}.
\end{eqnarray}
Both the general coordinate and $U(1)$ gauge transformations affect
the gauge field, shown above. Write the above transformations in the
form of $\delta\varphi^{i}=K_{\alpha}^{i}\delta\epsilon^{\alpha}$
where $\varphi^{i}=\left(g_{\mu\nu},\: A_{\mu}\right)$ and $\epsilon^{i}=\left(\epsilon^{\mu},\epsilon\right)$,
explicitly, we have
\begin{eqnarray}
\delta g_{\mu\nu}(x) & = & \int d^{4}x'\left[K^{g_{\mu\nu}(x)}{}_{\rho}(x,x')\delta\epsilon^{\rho}(x')+K^{g_{\mu\nu}(x)}(x,x')\delta\epsilon(x')\right],\\
\delta A_{\mu}(x) & = & \int d^{4}x'\left[K^{A_{\mu}(x)}{}_{\rho}(x,x')\delta\epsilon^{\rho}(x')+K^{A_{\mu}(x)}(x,x')\delta\epsilon(x')\right].
\end{eqnarray}
 The generators $K_{\alpha}^{i}$ for symmetric transformations defined
above are given by
\begin{eqnarray}
K^{g_{\mu\nu}(x)}{}_{\rho}(x,x') & = & \left[-g_{\mu\nu,\rho}(x)-2g_{\rho(\mu}(x)\partial_{\nu)}\right]\delta(x,x')\label{KG1}\\
K^{g_{\mu\nu}(x)}(x,x') & = & 0\label{KG2}\\
K^{A_{\mu}(x)}{}_{\rho}(x,x') & = & \left[-A_{\mu,\rho}(x)-A_{\rho}(x)\partial_{\mu}\right]\delta(x,x')\label{KA1}\\
K^{A_{\mu}(x)}(x,x') & = & \partial_{\mu}\delta(x,x')\label{KA2}
\end{eqnarray}
 The parentheses mean the symmetrization over enclosed indices. $\delta(x,x')$
has the following features
\begin{eqnarray}
\int d^{4}x'F(x')\delta(x,x') & = & F(x)\nonumber \\
\int d^{4}x'F(x')\partial_{\mu}\delta(x,x') & = & \partial_{\mu}F(x)\nonumber \\
\int d^{4}x'F(x')\partial_{\mu}\partial_{\nu}\delta(x,x') & = & \partial_{\mu}\partial_{\nu}F(x)
\end{eqnarray}
 The explicit form of $\delta(x,x')$ is not important here, all we
need in the calculation are the features above. It can be shown that $\delta(x,x')$
of the following form can satisfy the above features
\begin{eqnarray*}
\delta(x,x') & = & |g(x')|^{\frac{1}{2}}\delta(x-x')|g(x)|^{-\frac{1}{2}}\textrm{ or simply }\delta(x-x')
\end{eqnarray*}
 where $\delta(x-x')$ is the usual Dirac $\delta$-function in flat
space-time. We will not rely on this explicit form of $\delta(x,x')$
in the calculations.

Now we should choose a proper metric on the field space. At first
sight, the metric seems arbitrary. It is suggested in \cite{Vilkovisky,BV}
that there are several guidelines or rules for the choice of the metric
being unique, the effects of metric have been discussed in \cite{Odintsov}
when relaxing one of the rules. The metric on the field space $\varphi^{i}$,
$G_{ij}$, can be defined by the line element, $ds^{2}=G_{ij}d\varphi^{i}d\varphi^{j}$,
\begin{equation}
ds^{2}=\int d^{n}xd^{n}x'\left\lbrace G_{g_{\mu\nu}(x)g_{\rho\sigma}(x')}dg_{\mu\nu}(x)dg_{\rho\sigma}(x')+G_{A_{\mu}(x)A_{\nu}(x')}dA_{\mu}(x)dA_{\nu}(x')\right\rbrace \label{Fmetric}
\end{equation}
 where the metric has the following form \cite{DJToms}
\begin{eqnarray}
G_{g_{\mu\nu}(x)g_{\rho\sigma}(x')} & = & \frac{1}{\kappa^{2}}|g(x)|^{\frac{1}{2}}\left(g^{\mu(\rho}g^{\sigma)\nu}-\frac{1}{2}g^{\mu\nu}g^{\rho\sigma}\right)\delta(x,x')\label{eq:GravityMetric}\\
G_{A_{\mu}(x)A_{\nu}(x')} & = & |g(x)|^{\frac{1}{2}}g^{\mu\nu}(x)\delta(x,x')\label{eq:GaugeMetric}
\end{eqnarray}
 The inverse metric is
\begin{eqnarray}
G^{g_{\mu\nu}(x)g_{\rho\sigma}(x')} & = & \kappa^{2}|g(x)|^{-\frac{1}{2}}\left(g_{\mu(\rho}g_{\sigma)\nu}-\frac{1}{2}g_{\mu\nu}g_{\rho\sigma}\right)\delta(x,x').\\
G^{A_{\mu}(x)A_{\nu}(x')} & = & |g(x)|^{-\frac{1}{2}}g_{\mu\nu}(x)\delta(x,x').
\end{eqnarray}
 The orthogonal relation $G^{ij}G_{jk}=\delta_{k}^{i}$ reads explicitly
as
\begin{eqnarray}
\int d^{4}x'G^{g_{\mu\nu}(x)g_{\rho\sigma}(x')}G_{g_{\rho\sigma}(x')g_{\lambda\tau}(x'')} & = & \delta_{(\mu}^{\lambda}\delta_{\nu)}^{\tau}\delta(x,x'')\\
\int d^{4}x'G^{A_{\mu}(x)A_{\nu}(x')}G_{A_{\nu}(x')A_{\rho}(x'')} & = & \delta_{\mu}^{\rho}\delta(x,x'')
\end{eqnarray}
 Using the metric and inverse metric on the field space, we can determine
the corresponding Christoffel connection through
\begin{equation}
\Gamma_{ij}^{k}=\frac{1}{2}G^{kl}\left[\frac{\delta G_{il}}{\delta\varphi^{j}}+\frac{\delta G_{jl}}{\delta\varphi^{i}}-\frac{\delta G_{ij}}{\delta\varphi^{l}}\right]
\end{equation}
Below, we show the details of the tedious calculation for the $\Gamma_{ij}^{k}$.

\section{Christoffel Connection on the field space}

In this section, we present the details to calculate the Christoffel
Connection on the field space \cite{DJToms}. Some useful formula
for derivation are listed below
\begin{eqnarray}
\frac{\delta g_{\mu\nu}(x)}{\delta g_{\rho\sigma}(x')} & = & \delta_{(\mu}^{\rho}\delta_{\nu)}^{\sigma}\delta(x,x'),\;\delta_{(\mu}^{\rho}\delta_{\nu)}^{\sigma}=\frac{1}{2}\left[\delta_{\mu}^{\rho}\delta_{\nu}^{\sigma}+\delta_{\nu}^{\rho}\delta_{\mu}^{\sigma}\right]\\
\frac{\delta A_{\mu}(x)}{\delta A_{\nu}(x')} & = & \delta_{\mu}^{\nu}\delta(x,x'),\;\frac{\delta g_{\mu\nu}(x)}{\delta A_{\rho}(x')}=0=\frac{\delta A_{\rho}(x)}{\delta g_{\mu\nu}(x')},\;\frac{\delta[\delta(x,x')]}{\delta\varphi^{i}}=0\\
\delta g^{\mu\nu}(x) & = & -g^{\mu\rho}(x)g^{\nu\sigma}(x)\delta g_{\rho\sigma}(x),\;\frac{\delta g^{\mu\nu}(x)}{\delta g_{\rho\sigma}(x')}=-g^{\mu(\rho}g^{\sigma)\nu}\delta(x,x')\\
\delta|g(x)|^{\frac{1}{2}} & = & \frac{1}{2}|g(x)|^{\frac{1}{2}}g^{\rho\sigma}(x)\delta g_{\rho\sigma}(x),\;\frac{\delta|g(x)|^{\frac{1}{2}}}{\delta g_{\rho\sigma}(x')}=\frac{1}{2}|g(x)|^{\frac{1}{2}}g^{\rho\sigma}(x)\delta(x,x')\label{eq:Identities}
\end{eqnarray}
 We can calculate the first non-zero component of Christoffel connection,
\begin{eqnarray}
 &  & \Gamma_{A_{\lambda}(x')A_{\tau}(x'')}^{g_{\mu\nu}(x)}=\int d^{4}\bar{x}\frac{1}{2}G^{g_{\mu\nu}(x)g_{\rho\sigma}(\bar{x})}\left[-\frac{\delta G_{A_{\lambda}(x')A_{\tau}(x'')}}{\delta g_{\rho\sigma}(\bar{x})}\right]\nonumber \\
 & = & \frac{1}{2}\kappa^{2}\delta_{(\mu}^{\lambda}\delta_{\nu)}^{\tau}\delta(x,x')\delta(x',x'')
\end{eqnarray}
 Note that there is a missing $\kappa^{2}$ in the corresponding equation
in \cite{DJToms}, but the final expanded action there includes the
$\kappa^{2}$ back. In deriving the above equation, we have used $G^{g_{\mu\nu}(x)A_{\rho}(\bar{x})}=0$,
$G_{g_{\mu\nu}(x)g_{\rho\sigma}(\bar{x}),A_{\tau}(x')}=0$ and
\begin{eqnarray}
\frac{\delta\left(|g(x')|^{\frac{1}{2}}g^{\lambda\tau}(x')\right)}{\delta g_{\rho\sigma}(\bar{x})} & = & -|g(x')|^{\frac{1}{2}}\left[g^{\lambda(\rho}g^{\sigma)\tau}-\frac{1}{2}g^{\rho\sigma}g^{\lambda\tau}\right]\delta(x',\bar{x})
\end{eqnarray}
 The next non-vanishing component is
\begin{eqnarray}
 &  & \Gamma_{A_{\nu}(x')g_{\alpha\beta}(x'')}^{A_{\mu}(x)}=\int d^{4}\bar{x}\frac{1}{2}G^{A_{\mu}(x)A_{\lambda}(\bar{x})}\left[\frac{\delta G_{A_{\nu}(x')A_{\lambda}(\bar{x})}}{\delta g_{\alpha\beta}(x'')}\right]\nonumber \\
 & = & \frac{1}{4}\left[g^{\alpha\beta}\delta_{\mu}^{\nu}-2g^{\nu(\alpha}\delta_{\mu}^{\beta)}\right]\delta(x,x')\delta(x',x'')=\Gamma_{g_{\alpha\beta}(x'')A_{\nu}(x')}^{A_{\mu}(x)}
\end{eqnarray}
 and the most complicated component is
\begin{eqnarray}
 &  & \Gamma_{g_{\mu\nu}(x')g_{\rho\sigma}(x'')}^{g_{\lambda\tau}(x)}\nonumber \\
 & = & \int d^{4}\bar{x}\frac{1}{2}G^{g_{\lambda\tau}(x)g_{\alpha\beta}(\bar{x})}\Biggl[\frac{\delta G_{g_{\mu\nu}(x')g_{\alpha\beta}(\bar{x})}}{\delta g_{\rho\sigma}(x'')}+\frac{\delta G_{g_{\alpha\beta}(\bar{x})g_{\rho\sigma}(x'')}}{\delta g_{\mu\nu}(x')}-\frac{\delta G_{g_{\mu\nu}(x')g_{\rho\sigma}(x'')}}{\delta g_{\alpha\beta}(\bar{x})}\Biggl]\label{eq:3GravityChn}
\end{eqnarray}
 This quantity is well-known in the literature, for instance \cite{FradkinTseytlin},
here we show the details as a check of our calculation. To calculate
the above expression, using eq.(\ref{eq:GravityMetric}) and eq.(\ref{eq:Identities}),
we can work out the first term in the bracket
\begin{eqnarray*}
\frac{\delta G_{g_{\mu\nu}(x')g_{\rho\sigma}(x'')}}{\delta g_{\alpha\beta}(\bar{x})} & = & -\frac{1}{2\kappa^{2}}\delta(x',x'')\delta(x',\bar{x})|g(x')|^{\frac{1}{2}}\left(\left[g^{\mu(\alpha}g^{\beta)\rho}-\frac{1}{2}g^{\alpha\beta}g^{\mu\rho}\right]g^{\nu\sigma}+g^{\mu\rho}g^{\nu(\alpha}g^{\beta)\sigma}\right)\\
 &  & -\frac{1}{2\kappa^{2}}\delta(x',x'')\delta(x',\bar{x})|g(x')|^{\frac{1}{2}}\left(\left[g^{\mu(\alpha}g^{\beta)\sigma}-\frac{1}{2}g^{\alpha\beta}g^{\mu\sigma}\right]g^{\nu\rho}+g^{\mu\sigma}g^{\nu(\alpha}g^{\beta)\rho}\right)\\
 &  & +\frac{1}{2\kappa^{2}}\delta(x',x'')\delta(x',\bar{x})|g(x')|^{\frac{1}{2}}\left(\left[g^{\mu(\alpha}g^{\beta)\nu}-\frac{1}{2}g^{\alpha\beta}g^{\mu\nu}\right]g^{\rho\sigma}+g^{\mu\nu}g^{\rho(\alpha}g^{\beta)\sigma}\right)
\end{eqnarray*}
 which can be rewritten symmetrically
\begin{eqnarray*}
\frac{\delta G_{g_{\mu\nu}(x')g_{\rho\sigma}(x'')}}{\delta g_{\alpha\beta}(\bar{x})} & = & \frac{1}{2\kappa^{2}}\delta(x',x'')\delta(x',\bar{x})|g(x')|^{\frac{1}{2}}\biggl[-\frac{1}{2}g^{\mu\nu}g^{\rho\sigma}g^{\alpha\beta}+g^{\alpha\beta}g^{\mu(\rho}g^{\sigma)\nu}+g^{\mu\nu}g^{\rho(\alpha}g^{\beta)\sigma}\\
 &  & {}+g^{\rho\sigma}g^{\mu(\alpha}g^{\beta)\nu}-g^{\mu\rho}g^{\nu(\alpha}g^{\beta)\sigma}-g^{\mu\sigma}g^{\nu(\alpha}g^{\beta)\rho}-g^{\nu\rho}g^{\mu(\alpha}g^{\beta)\sigma}-g^{\nu\sigma}g^{\mu(\alpha}g^{\beta)\rho}\biggl]
\end{eqnarray*}
 We can see that the tensor structure is symmetric under $\mu\leftrightarrow\nu$,
$\rho\leftrightarrow\sigma$, $\alpha\leftrightarrow\beta$ and $(\mu\nu)\leftrightarrow(\rho\sigma)$,
as expected since we have the symmetric metric $G_{ij}=G_{ji}$. The
next two terms in the bracket of eq.(\ref{eq:3GravityChn}) can be directly
written down
\begin{eqnarray*}
\frac{\delta G_{g_{\mu\nu}(x')g_{\alpha\beta}(\bar{x})}}{\delta g_{\rho\sigma}(x'')} & = & \frac{1}{2\kappa^{2}}\delta(x',\bar{x})\delta(x',x'')|g(x')|^{\frac{1}{2}}\biggl[-\frac{1}{2}g^{\mu\nu}g^{\rho\sigma}g^{\alpha\beta}+g^{\rho\sigma}g^{\mu(\alpha}g^{\beta)\nu}+g^{\mu\nu}g^{\alpha(\rho}g^{\sigma)\beta}\\
 &  & {}+g^{\alpha\beta}g^{\mu(\rho}g^{\sigma)\nu}-g^{\mu\alpha}g^{\nu(\rho}g^{\sigma)\beta}-g^{\mu\beta}g^{\nu(\rho}g^{\sigma)\alpha}-g^{\nu\alpha}g^{\mu(\rho}g^{\sigma)\beta}-g^{\nu\beta}g^{\mu(\rho}g^{\sigma)\alpha}\biggl]
\end{eqnarray*}
\begin{eqnarray*}
\frac{\delta G_{g_{\alpha\beta}(\bar{x})g_{\rho\sigma}(x'')}}{\delta g_{\mu\nu}(x')} & = & \frac{1}{2\kappa^{2}}\delta(\bar{x},x'')\delta(\bar{x},x')|g(\bar{x})|^{\frac{1}{2}}\biggl[-\frac{1}{2}g^{\mu\nu}g^{\rho\sigma}g^{\alpha\beta}+g^{\mu\nu}g^{\alpha(\rho}g^{\sigma)\beta}+g^{\rho\sigma}g^{\alpha(\mu}g^{\nu)\beta}\\
 &  & {}+g^{\alpha\beta}g^{\rho(\mu}g^{\nu)\sigma}-g^{\alpha\rho}g^{\beta(\mu}g^{\nu)\sigma}-g^{\alpha\sigma}g^{\beta(\mu}g^{\nu)\rho}-g^{\beta\rho}g^{\alpha(\mu}g^{\nu)\sigma}-g^{\beta\sigma}g^{\alpha(\mu}g^{\nu)\rho}\biggl]
\end{eqnarray*}
 These three components have a common factor,
\begin{equation}
\left[-\frac{1}{2}g^{\mu\nu}g^{\rho\sigma}g^{\alpha\beta}+g^{\rho\sigma}g^{\mu(\alpha}g^{\beta)\nu}+g^{\mu\nu}g^{\alpha(\rho}g^{\sigma)\beta}+g^{\alpha\beta}g^{\mu(\rho}g^{\sigma)\nu}\right]
\end{equation}
 which is symmetric under $(\mu\nu)\leftrightarrow(\rho\sigma)\leftrightarrow(\alpha\beta)$.
Now eq.(\ref{eq:3GravityChn}) can be calculated as
\begin{eqnarray}
 &  & \Gamma_{g_{\mu\nu}(x')g_{\rho\sigma}(x'')}^{g_{\lambda\tau}(x)}\nonumber \\
 & = & \int d^{4}\bar{x}\frac{1}{2}G^{g_{\lambda\tau}(x)g_{\alpha\beta}(\bar{x})}\left[\frac{\delta G_{g_{\mu\nu}(x')g_{\alpha\beta}(\bar{x})}}{\delta g_{\rho\sigma}(x'')}+\frac{\delta G_{g_{\alpha\beta}(\bar{x})g_{\rho\sigma}(x'')}}{\delta g_{\mu\nu}(x')}-\frac{\delta G_{g_{\mu\nu}(x')g_{\rho\sigma}(x'')}}{\delta g_{\alpha\beta}(\bar{x})}\right]\nonumber \\
 & = & \frac{1}{4}\delta(x,x')\delta(x',x'')\left(g_{\lambda(\alpha}g_{\beta)\tau}-\frac{1}{2}g_{\lambda\tau}g_{\alpha\beta}\right)\times\nonumber \\
 &  & \biggl(\left[-\frac{1}{2}g^{\mu\nu}g^{\rho\sigma}g^{\alpha\beta}+g^{\rho\sigma}g^{\mu(\alpha}g^{\beta)\nu}+g^{\mu\nu}g^{\alpha(\rho}g^{\sigma)\beta}+g^{\alpha\beta}g^{\mu(\rho}g^{\sigma)\nu}\right]\nonumber \\
 &  & +\left[-g^{\mu\alpha}g^{\nu(\rho}g^{\sigma)\beta}-g^{\mu\beta}g^{\nu(\rho}g^{\sigma)\alpha}-g^{\nu\alpha}g^{\mu(\rho}g^{\sigma)\beta}-g^{\nu\beta}g^{\mu(\rho}g^{\sigma)\alpha}\right]\nonumber \\
 &  & +\left[-g^{\alpha\rho}g^{\beta(\mu}g^{\nu)\sigma}-g^{\alpha\sigma}g^{\beta(\mu}g^{\nu)\rho}-g^{\beta\rho}g^{\alpha(\mu}g^{\nu)\sigma}-g^{\beta\sigma}g^{\alpha(\mu}g^{\nu)\rho}\right]\nonumber \\
 &  & -\left[-g^{\mu\rho}g^{\nu(\alpha}g^{\beta)\sigma}-g^{\mu\sigma}g^{\nu(\alpha}g^{\beta)\rho}-g^{\nu\rho}g^{\mu(\alpha}g^{\beta)\sigma}-g^{\nu\sigma}g^{\mu(\alpha}g^{\beta)\rho}\right]\biggl)\label{eq:3GravityChn1}
\end{eqnarray}
 The index contract can be computed directly. We finally have
\begin{eqnarray}
\Gamma_{g_{\mu\nu}(x')g_{\rho\sigma}(x'')}^{g_{\lambda\tau}(x)} & = & \delta(x,x')\delta(x',x'')\biggl(-\frac{1}{8}g^{\mu\nu}g^{\rho\sigma}g_{\lambda\tau}-\delta_{(\lambda}^{(\mu}g^{\nu)(\rho}\delta_{\tau)}^{\sigma)}\nonumber \\
 &  & +\frac{1}{4}\left[g^{\rho\sigma}\delta_{(\lambda}^{\mu}\delta_{\tau)}^{\nu}+g^{\mu\nu}\delta_{(\lambda}^{\rho}g_{\tau)}^{\sigma}+g_{\lambda\tau}g^{\mu(\rho}g^{\sigma)\nu}\right]\biggl)
\end{eqnarray}
We can summarize the non-vanishing Christoffel connection components
as follows
\begin{eqnarray}
\Gamma_{A_{\lambda}(x')A_{\tau}(x'')}^{g_{\mu\nu}(x)} & = & \frac{1}{2}\kappa^{2}\delta_{\mu}^{(\lambda}\delta_{\nu}^{\tau)}\delta(x,x')\delta(x',x'')\nonumber \\
\Gamma_{A_{\nu}(x')g_{\alpha\beta}(x'')}^{A_{\mu}(x)} & = & \frac{1}{4}\left(\delta_{\mu}^{\nu}g^{\alpha\beta}-2\delta_{\mu}^{(\alpha}g^{\beta)\nu}\right)\delta(x,x')\delta(x,x'')\nonumber \\
\Gamma_{g_{\alpha\beta}(x'')A_{\nu}(x')}^{A_{\mu}(x)} & = & \Gamma_{A_{\nu}(x')g_{\alpha\beta}(x'')}^{A_{\mu}(x)}\nonumber \\
\Gamma_{g_{\mu\nu}(x')g_{\rho\sigma}(x'')}^{g_{\lambda\tau}(x)} & = & \biggl[-\delta_{(\lambda}^{(\mu}g^{\nu)(\rho}\delta_{\tau)}^{\sigma)}+\frac{1}{4}g^{\mu\nu}\delta_{(\lambda}^{\rho}\delta_{\tau)}^{\sigma}+\frac{1}{4}g^{\rho\sigma}\delta_{(\lambda}^{\mu}\delta_{\tau)}^{\nu}\nonumber \\
 &  & +\frac{1}{4}\left(g_{\lambda\tau}g^{\mu(\rho}g^{\sigma)\nu}-\frac{1}{2}g_{\lambda\tau}g^{\mu\nu}g^{\rho\sigma}\right)\biggl]\delta(x,x'')\delta(x',x'')\label{eq:connection}
\end{eqnarray}

\section{Functional derivatives}

Let us now calculate the functional derivatives $S_{,i}$ and $S_{,ij}$. The
functional derivatives over the graviton and gauge fields on a general
background are decomposed as
\begin{eqnarray}
\frac{\delta S}{\delta g_{\mu\nu}(x)} & = & \frac{\delta\left(S_{G}+S_{M}\right)}{\delta g_{\mu\nu}(x)}=\frac{\delta S_{G}}{\delta g_{\mu\nu}(x)}+\frac{\delta S_{M}}{\delta g_{\mu\nu}(x)}\\
\frac{\delta S}{\delta A_{\mu}(x)} & = & \frac{\delta\left(S_{G}+S_{M}\right)}{\delta A_{\mu}(x)}=\frac{\delta S_{M}}{\delta A_{\mu}(x)}
\end{eqnarray}
 We shall calculate the above quantities separately.
\begin{eqnarray}
\frac{\delta S_{M}}{\delta A_{\mu}(x)} & = & \frac{1}{4}\int d^{4}x'|g(x')|^{\frac{1}{2}}g^{\alpha\rho}g^{\beta\sigma}\frac{\delta\left(F_{\rho\sigma}F_{\alpha\beta}\right)}{\delta A_{\mu}(x)}=\partial_{\alpha}\left(|g(x)|^{\frac{1}{2}}F^{\mu\alpha}\right)
\end{eqnarray}
 and
\begin{eqnarray}
 &  & \frac{\delta S_{M}}{\delta g_{\mu\nu}(x)}=\frac{1}{4}\int d^{4}x'F_{\rho\sigma}F_{\alpha\beta}\frac{\delta\left(|g(x')|^{1/2}g^{\alpha\rho}g^{\beta\sigma}\right)}{\delta g_{\mu\nu}(x)}\nonumber \\
 & = & \frac{1}{4}|g(x)|^{\frac{1}{2}}\left[\frac{1}{2}g^{\mu\nu}F^{2}-2F^{\mu}{}_{\sigma}F^{\nu\sigma}\right]=-\frac{1}{2}|g(x)|^{\frac{1}{2}}T^{\mu\nu}
\end{eqnarray}
 where we have used $F^{2}=F_{\alpha\beta}F^{\alpha\beta}$ for short and
defined
\begin{equation}
T^{\mu\nu}=\frac{-2}{\sqrt{g(x)}}\frac{\delta S_{M}}{\delta g_{\mu\nu}(x)}=F^{\mu}{}_{\sigma}F^{\nu\sigma}-\frac{1}{4}g^{\mu\nu}F^{2}
\end{equation}
 For the functional derivative with respect to the metric, we list
the following formulas for convenience.
\begin{eqnarray}
\delta R^{\rho}{}_{\sigma\mu\nu} & = & \partial_{\mu}\delta\Gamma_{\nu\sigma}^{\rho}-\partial_{\nu}\delta\Gamma_{\mu\sigma}^{\rho}+\delta\left[\Gamma_{\mu\lambda}^{\rho}\Gamma_{\nu\sigma}^{\lambda}-\Gamma_{\mu\lambda}^{\rho}\Gamma_{\nu\sigma}^{\lambda}\right]\nonumber \\
\nabla_{\lambda}\delta\Gamma_{\nu\mu}^{\rho} & = & \partial_{\lambda}\delta\Gamma_{\nu\mu}^{\rho}+\Gamma_{\sigma\lambda}^{\rho}\delta\Gamma_{\nu\mu}^{\sigma}-\Gamma_{\nu\lambda}^{\sigma}\delta\Gamma_{\sigma\mu}^{\rho}-\Gamma_{\mu\lambda}^{\sigma}\delta\Gamma_{\nu\sigma}^{\rho}\nonumber \\
\delta R^{\rho}{}_{\sigma\mu\nu} & = & \nabla_{\mu}\delta\Gamma_{\nu\sigma}^{\rho}-\nabla_{\nu}\delta\Gamma_{\mu\sigma}^{\rho}\nonumber \\
\delta\Gamma_{\mu\nu}^{\rho} & = & \frac{1}{2}g^{\rho\sigma}\left[\left(\delta g_{\mu\sigma}\right)_{;\nu}+\left(\delta g_{\nu\sigma}\right)_{;\mu}-\left(\delta g_{\mu\nu}\right)_{;\rho}\right]\label{eq:variationR}
\end{eqnarray}
 and
\begin{eqnarray*}
\delta R_{\mu\nu} & = & \delta R^{\rho}{}_{\mu\rho\nu}=\nabla_{\rho}\delta\Gamma_{\nu\mu}^{\rho}-\nabla_{\nu}\delta\Gamma_{\rho\mu}^{\rho}\\
\delta R & = & R_{\mu\nu}\delta g^{\mu\nu}+g^{\mu\nu}\delta R_{\mu\nu}=R_{\mu\nu}\delta g^{\mu\nu}+\nabla_{\sigma}\left[g^{\mu\nu}\delta\Gamma_{\nu\mu}^{\sigma}-g^{\mu\sigma}\delta\Gamma_{\rho\mu}^{\rho}\right]\\
 & = & R_{\mu\nu}\delta g^{\mu\nu}+g^{\mu\nu}g^{\rho\sigma}\left[\delta g_{\rho\sigma;\mu\nu}+\delta g_{\rho\mu;\rho\nu}\right]
\end{eqnarray*}
 Similarly, we have
\begin{eqnarray*}
\frac{\delta S_{G}}{\delta g_{\mu\nu}(x)} & = & =-\frac{2}{\kappa^{2}}\int d^{4}x'\left[\frac{\delta\left(|g(x')|^{\frac{1}{2}}(R-2\Lambda)\right)}{\delta g_{\mu\nu}(x)}\right]=-\frac{2}{\kappa^{2}}|g(x)|^{\frac{1}{2}}E^{\mu\nu}
\end{eqnarray*}
where we have used
\begin{eqnarray}
\int d^{4}x'|g(x')|^{\frac{1}{2}}\left(g^{\alpha\beta}\frac{\delta R_{\alpha\beta}}{\delta g_{\mu\nu}(x)}\right) & = & \textrm{Surface terms}
\end{eqnarray}
 and defined $E^{\mu\nu}=\frac{1}{2}(R-2\Lambda)g^{\mu\nu}-R^{\mu\nu}$.
The Einstein equation can be obtained by imposing $\frac{\delta S}{\delta g_{\mu\nu}(x)}=0$,
\begin{equation}
R^{\mu\nu}-\frac{1}{2}Rg^{\mu\nu}+\Lambda g^{\mu\nu}=8\pi GT^{\mu\nu}
\end{equation}
 In the present paper, we are working in a flat background space-time. Since in
this case we expand at a background that doesn't satisfy Einstein
equation, we actually deal with the off-shell effective action. The
connection term in eq.(\ref{Eq:VDaction}) is necessary
to be included for a gauge condition independent result. For $S_{,i}$
, we summarize the final result as
\begin{eqnarray}
\frac{\delta S_{M}}{\delta g_{\mu\nu}(x)} & = & \frac{1}{4}|g(x)|^{\frac{1}{2}}\left[\frac{1}{2}g^{\mu\nu}F_{\alpha\beta}F^{\alpha\beta}-2F^{\mu}{}_{\sigma}F^{\nu\sigma}\right]\\
\frac{\delta S_{G}}{\delta g_{\mu\nu}(x)} & = & -\frac{2}{\kappa^{2}}|g(x)|^{\frac{1}{2}}\left[\frac{1}{2}(R-2\Lambda)g^{\mu\nu}-R^{\mu\nu}\right]\\
\frac{\delta S_{M}}{\delta A_{\mu}(x)} & = & \partial_{\alpha}\left(|g(x)|^{\frac{1}{2}}F^{\mu\alpha}\right),\;\frac{\delta S_{G}}{\delta A_{\mu}(x)}=0
\end{eqnarray}
 The above formulas are true for general background space-time $\bar{g}_{\mu\nu}$.

Although we can expand the Lagrangian eq.(\ref{eq:GravityL})
straightforwardly for a flat background space-time, here we shall
calculate $S_{,ij}$ for the expansion. We may use the following
equations for convenience,
\begin{eqnarray*}
 &  & \frac{\delta\left(|g(x')|^{\frac{1}{2}}F^{2}\right)}{\delta g_{\mu\nu}(x)}=|g(x)|^{\frac{1}{2}}\delta(x',x)\left[\frac{1}{2}g^{\mu\nu}F^{2}-2F^{\mu}{}_{\sigma}F^{\nu\sigma}\right]\\
 &  & \frac{\delta\left(|g(x')|^{\frac{1}{2}}F^{\alpha\beta}\right)}{\delta g_{\mu\nu}(x)}=\frac{\delta\left(|g(x')|^{\frac{1}{2}}g^{\alpha\rho}g^{\beta\sigma}F_{\rho\sigma}\right)}{\delta g_{\mu\nu}(x)}\\
 & = & |g(x')|^{\frac{1}{2}}\delta(x',x)\left[\frac{1}{2}g^{\mu\nu}F^{\alpha\beta}-g^{\alpha(\mu}F^{\nu)\beta}+g^{\beta(\mu}F^{\nu)\alpha}\right]
\end{eqnarray*}
In computing $S_{,ij}$, the following components are straightforward,
\begin{eqnarray}
\frac{\delta^{2}S_{G}}{\delta A_{\mu}(x)\delta A_{\nu}(x')} & = & 0\\
\frac{\delta^{2}S_{M}}{\delta A_{\mu}(x)\delta A_{\nu}(x')} & = & \partial_{\alpha}\left(|g(x)|^{\frac{1}{2}}\frac{\delta F^{\mu\alpha}}{\delta A_{\nu}(x')}\right)\nonumber \\
 & = & \partial_{\alpha}\left(|g(x)|^{\frac{1}{2}}\left[\partial^{\mu}\delta^{\alpha\nu}-\partial^{\alpha}\delta^{\mu\nu}\right]\delta(x,x')\right)\\
\frac{\delta^{2}S_{M}}{\delta A_{\mu}(x)\delta g_{\alpha\beta}(x')} & = & \partial_{\nu}\left(\frac{\delta\left[|g(x)|^{\frac{1}{2}}F^{\mu\nu}\right]}{\delta g_{\alpha\beta}(x')}\right)\nonumber \\
 & = & \partial_{\nu}\left(g(x)|^{\frac{1}{2}}\delta(x,x')\left[\frac{1}{2}g^{\alpha\beta}F^{\mu\nu}-g^{\mu(\alpha}F^{\beta)\nu}+g^{\nu(\alpha}F^{\beta)\mu}\right]\right)
\end{eqnarray}
The rest parts are much more complicated. The matter part has the form
\begin{eqnarray*}
 &  & \frac{\delta^{2}S_{M}}{\delta g_{\mu\nu}(x)\delta g_{\alpha\beta}(x')}=\frac{1}{4}\frac{\delta}{\delta g_{\alpha\beta}(x')}\left(|g(x)|^{\frac{1}{2}}\left[\frac{1}{2}g^{\mu\nu}F^{2}-2F^{\mu}{}_{\sigma}F^{\nu\sigma}\right]\right)\\
 & = & |g(x)|^{\frac{1}{2}}\delta(x,x')\Biggl(\frac{1}{16}g^{\mu\nu}g^{\alpha\beta}F^{2}-\frac{1}{8}F^{2}g^{\mu(\alpha}g^{\beta)\nu}+\frac{1}{2}F^{\nu}{}_{\sigma}g^{\mu(\alpha}F^{\beta)\sigma}\\
 &  & +\frac{1}{2}F^{\mu}{}_{\sigma}g^{\nu(\alpha}F^{\beta)\sigma}-\frac{1}{2}F^{\mu}{}_{\sigma}g^{\sigma(\alpha}F^{\beta)\nu}-\frac{1}{4}g^{\mu\nu}F^{\alpha}{}_{\sigma}F^{\beta\sigma}-\frac{1}{4}g^{\alpha\beta}F^{\mu}{}_{\sigma}F^{\nu\sigma}\Biggl)
\end{eqnarray*}
For the gravity part, we have
\begin{eqnarray*}
 &  & \frac{\delta^{2}S_{G}}{\delta g_{\mu\nu}(x)\delta g_{\alpha\beta}(x')}\\
 & = & -\frac{1}{\kappa^{2}}|g(x)|^{\frac{1}{2}}\delta(x,x')\left((R-2\Lambda)\left[\frac{1}{2}g^{\mu\nu}g^{\alpha\beta}-g^{\mu(\alpha}g^{\beta)\nu}\right]-g^{\mu\nu}R^{\alpha\beta}-g^{\alpha\beta}R^{\mu\nu}\right.\\
 &  & +2\left[g^{\alpha(\mu}R^{\beta)\nu}+g^{\alpha(\nu}R^{\beta)\mu}\right]\biggl)-\frac{1}{\kappa^{2}}|g(x)|^{\frac{1}{2}}\left[g^{\mu\nu}g^{\rho\sigma}-2g^{\mu\rho}g^{\sigma\nu}\right]\frac{\delta R_{\rho\sigma}}{\delta g_{\alpha\beta}(x')}
\end{eqnarray*}
 where $\frac{\delta R_{\rho\sigma}}{\delta g_{\alpha\beta}(x')}$ can be
worked out by using eq.(\ref{eq:variationR}).

\section{Euclidean flat Background}

In the following discussion, we will focus on the flat background
space-time and consider the one-loop contribution to the gauge effective
action from the graviton. We expand the fields, $\varphi^{i}=(g_{\mu\nu},A_{\mu})$,
at the background-fields, $\bar{\varphi}^{i}=(\delta_{\mu\nu},\bar{A}_{\mu})$,
\begin{eqnarray}
g_{\mu\nu} & = & \delta_{\mu\nu}+\kappa h_{\mu\nu};\quad A_{\mu}=\bar{A}_{\mu}+a_{\mu}
\end{eqnarray}
Expansion of the action in flat background space-time is straightforward
by directly replacing the fields with above equations. One can also
work out first the functional derivatives, $S_{,i}$ and $S_{,ij}$,
and then consider the effective Lagrangian $\frac{1}{2}\eta^{i}\eta^{j}\left[S_{,ij}-\Gamma_{ij}^{k}S_{,k}+\frac{1}{2\Omega}\chi_{\alpha}{}^{,i}\chi^{\alpha}{}_{,j}\right]$,
as we shall show below in detail.

Using the formulas given in the previous appendix sections and imposing the flat background
space-time, we have
\begin{eqnarray}
\left.\frac{\delta S_{M}}{\delta g_{\mu\nu}(x)}\right|_{\bar{\varphi}^{i}} & = & \frac{1}{4}\left[\frac{1}{2}\delta^{\mu\nu}\bar{F}{}^{2}-2\bar{F}{}^{\mu}{}_{\sigma}\bar{F}{}^{\nu\sigma}\right],\;\left.\frac{\delta S_{G}}{\delta g_{\mu\nu}(x)}\right|_{\bar{\varphi}^{i}}=\frac{2}{\kappa^{2}}\Lambda\delta^{\rho\sigma}\label{eq:FirstDflat}\\
\left.\frac{\delta S_{M}}{\delta A_{\mu}(x)}\right|_{\bar{\varphi}^{i}} & = & \partial_{\alpha}\bar{F}{}^{\mu\alpha},\;\left.\frac{\delta S_{G}}{\delta A_{\mu}(x)}\right|_{\bar{\varphi}^{i}}=0
\end{eqnarray}
 For $S_{,ij}$, the most complicated one is $\frac{\delta^{2}S}{\delta g_{\mu\nu}(x)\delta g_{\alpha\beta}(x')}$,
and we need the following result
\begin{eqnarray*}
\left.\frac{\delta R_{\mu\nu}}{\delta g_{\alpha\beta}(x')}\right|_{\bar{\varphi}^{i}} & = & \frac{1}{2}\left[2\delta_{(\mu}^{(\alpha}\partial^{\beta)}\partial_{\nu)}-\delta_{\mu}^{(\alpha}\delta_{\nu}^{\beta)}\square-\delta^{\alpha\beta}\partial_{\mu}\partial_{\nu}\right]\delta(x,x')
\end{eqnarray*}
putting all together, we can show that
\begin{eqnarray*}
 &  & \left.\frac{\delta^{2}S_{G}}{\delta g_{\mu\nu}(x)\delta g_{\alpha\beta}(x')}\right|_{\bar{\varphi}^{i}}\\
 & = & \frac{1}{\kappa^{2}}\delta(x,x')\Bigl[\partial^{(\mu}\delta^{\nu)(\alpha}\partial^{\beta)}+\left(\delta^{\alpha\beta}\delta^{\mu\nu}-\delta^{\mu(\alpha}\delta^{\beta)\nu}\right)\square-\delta^{\mu\nu}\partial^{\alpha}\partial^{\beta}-\delta^{\alpha\beta}\partial^{\mu}\partial^{\nu}\Bigl]\\
 &  & +\frac{2\Lambda}{\kappa^{2}}\delta(x,x')\left[\frac{1}{2}\delta^{\mu\nu}\delta^{\alpha\beta}-\delta^{\mu(\alpha}\delta^{\beta)\nu}\right]
\end{eqnarray*}
We may summarize the following formulas with flat background space-time,
\begin{eqnarray*}
\left.\frac{\delta^{2}S_{G}}{\delta A_{\mu}(x)\delta A_{\nu}(x')}\right|_{\bar{\varphi}^{i}} & = & \left.\frac{\delta^{2}S_{G}}{\delta A_{\mu}(x)\delta g_{\alpha\beta}(x')}\right|_{\bar{\varphi}^{i}}=0\\
\left.\frac{\delta^{2}S_{M}}{\delta A_{\mu}(x)\delta A_{\nu}(x')}\right|_{\bar{\varphi}^{i}} & = & \left[\partial^{\mu}\partial^{\nu}-\partial^{2}\delta^{\mu\nu}\right]\delta(x,x')\\
\left.\frac{\delta^{2}S_{M}}{\delta A_{\mu}(x)\delta g_{\alpha\beta}(x')}\right|_{\bar{\varphi}^{i}} & = & \partial_{\nu}\left(\delta(x,x')\left[\frac{1}{2}\delta^{\alpha\beta}\bar{F}{}^{\mu\nu}-\delta^{\mu(\alpha}\bar{F}{}^{\beta)\nu}+\delta^{\nu(\alpha}\bar{F}{}^{\beta)\mu}\right]\right)
\end{eqnarray*}
 and
\begin{eqnarray*}
\left.\frac{\delta^{2}S_{M}}{\delta g_{\mu\nu}(x)\delta g_{\alpha\beta}(x')}\right|_{\bar{\varphi}^{i}} & = & \delta(x,x')\left(\left[\frac{1}{16}\delta^{\mu\nu}\delta^{\alpha\beta}\bar{F}{}^{2}-\frac{1}{8}\bar{F}{}^{2}\delta^{\mu(\alpha}\delta^{\beta)\nu}\right]\right.\\
 &  & +\frac{1}{2}\bar{F}{}^{\nu}{}_{\sigma}\delta^{\mu(\alpha}\bar{F}{}^{\beta)\sigma}+\frac{1}{2}\bar{F}{}^{\mu}{}_{\sigma}\delta^{\nu(\alpha}\bar{F}{}^{\beta)\sigma}-\frac{1}{2}\bar{F}{}^{\mu}{}_{\sigma}\delta^{\sigma(\alpha}\bar{F}{}^{\beta)\nu}\\
 &  & \left.-\frac{1}{4}\delta^{\mu\nu}\bar{F}{}^{\alpha}{}_{\sigma}\bar{F}{}^{\beta\sigma}-\frac{1}{4}\delta^{\alpha\beta}\bar{F}{}^{\mu}{}_{\sigma}\bar{F}{}^{\nu\sigma}\right)\\
\left.\frac{\delta^{2}S_{G}}{\delta g_{\mu\nu}(x)\delta g_{\alpha\beta}(x')}\right|_{\bar{\varphi}^{i}} & = & \frac{2\Lambda}{\kappa^{2}}\delta(x,x')\left[\frac{1}{2}\delta^{\mu\nu}\delta^{\alpha\beta}-\delta^{\mu(\alpha}\delta^{\beta)\nu}\right]+\frac{1}{\kappa^{2}}\delta(x,x')\times\\
 &  & \left[\partial^{(\mu}\delta^{\nu)(\alpha}\partial^{\beta)}+\left(\delta^{\alpha\beta}\delta^{\mu\nu}-\delta^{\mu(\alpha}\delta^{\beta)\nu}\right)\square-\delta^{\mu\nu}\partial^{\alpha}\partial^{\beta}-\delta^{\alpha\beta}\partial^{\mu}\partial^{\nu}\right]
\end{eqnarray*}
so far we have the pieces to calculate the covariant derivative for the classical
action with respect to $\varphi^{i}$ and to expand the terms which
are necessary for one-loop calculation of Vilkovisky-DeWitt effective
action. We shall work out the needed effective action by expanding and truncating piece by piece.

\subsection{Ordinary derivative terms $\frac{1}{2}\eta^{i}S_{,ij}\eta^{j}$}

Let us first consider the quadratic terms on the quantum gauge field $a_{\mu}$,
\begin{eqnarray}
\frac{1}{2}aS_{,AA}a & = & \frac{1}{2}\int d^{4}xd^{4}x'a_{\mu}(x)\left.\frac{\delta^{2}S_{M}}{\delta A_{\mu}(x)\delta A_{\nu}(x')}\right|_{\bar{\varphi}^{i}}a_{\nu}(x')\nonumber \\
 & = & \frac{1}{2}\int d^{4}xa_{\mu}(x)\left[\partial^{\mu}\partial^{\nu}-\partial^{2}\delta^{\mu\nu}\right]a_{\nu}(x)
\end{eqnarray}
 This part together with the gauge fixing term will give the propagator
of gauge boson. With including cosmological constant, it will be seen in the connection
terms that other terms will contribution as well. The quadratic terms
cross on graviton $h_{\mu\nu}$ and gauge field $a_{\mu}$ are given by 
\begin{eqnarray}
 &  & \frac{1}{2}aS_{,Ag}\kappa h=\frac{1}{2}\int d^{4}xd^{4}x'a_{\mu}(x)\left.\frac{\delta^{2}S_{M}}{\delta A_{\mu}(x)\delta g_{\alpha\beta}(x')}\right|_{\bar{\varphi}^{i}}\kappa h_{\alpha\beta}(x')\nonumber \\
 & = & \frac{\kappa}{2}\int d^{4}x\left[\frac{1}{2}h\bar{F}{}^{\mu\nu}\partial_{\mu}a_{\nu}-\bar{F}_{\beta}{}^{\mu}h^{\nu\beta}\partial_{\nu}a_{\mu}+\bar{F}_{\beta}{}^{\nu}h^{\alpha\beta}\partial_{\nu}a_{\alpha}\right]
\end{eqnarray}
 and the same terms for $\frac{1}{2}\kappa hS_{,gA}a$. There are
also quadratic terms on graviton field $h_{\mu\nu}$,
\begin{eqnarray}
 &  & \frac{1}{2}\kappa hS_{,gg}\kappa h=\frac{\kappa^{2}}{2}\int d^{4}xd^{4}x'h_{\mu\nu}(x)\left.\frac{\delta^{2}S_{G}}{\delta g_{\mu\nu}(x)\delta g_{\alpha\beta}(x')}\right|_{\bar{\varphi}^{i}}h_{\alpha\beta}(x')\nonumber \\
 & = & \int d^{4}x\left(\Lambda\left[\frac{1}{2}h^{2}-h^{\mu\nu}h_{\mu\nu}\right]+\left[\frac{1}{4}h\partial^{2}h-\frac{1}{2}h^{\mu\nu}\partial^{2}h_{\mu\nu}-\left(\partial^{\mu}h_{\mu\nu}-\frac{1}{2}\partial_{\nu}h\right)^{2}\right]\right)
\end{eqnarray}
 The first term associated with $\Lambda$ in the parenthesis may act as a mass term
for graviton, and will display itself in the graviton propagator.

\subsection{Manifold connection terms $-\frac{1}{2}\eta^{i}\Gamma_{ij}^{k}S_{,k}\eta^{j}$}

Since we are working in a flat background space-time, which is not
a solution of Einstein equation in the presence of matter fields,
we need to include the connection terms for yielding a gauge fixing condition
independent result in the Vilkovisky-DeWitt's framework. The quadratic
terms on quantum gauge field $a_{\mu}$ are,
\begin{eqnarray}
 &  & -\frac{1}{2}a\Gamma_{AA}^{g}aS_{,g}=-\frac{1}{2}\int d^{4}xd^{4}x'd^{4}x''a_{\lambda}(x')\Gamma_{A_{\lambda}(x')A_{\tau}(x'')}^{g_{\mu\nu}(x)}a_{\tau}(x'')\left.\frac{\delta S}{\delta g_{\mu\nu}(x)}\right|_{\bar{\varphi}^{i}}\nonumber \\
 & = & \int d^{4}x\left(-\frac{1}{2}\Lambda\delta^{\mu\nu}-\frac{1}{8}\kappa^{2}\left[\frac{1}{4}\delta^{\mu\nu}\bar{F}{}^{2}-\bar{F}{}^{\mu}{}_{\sigma}\bar{F}{}^{\nu\sigma}\right]\right)a_{\mu}a_{\nu}
\end{eqnarray}
As we can see from the above equation that, with the non-zero cosmological
constant, the connection induced interactions with the term $2\Lambda a_{\mu}a^{\nu}$
will change the gauge propagator. This also happens in the graviton
part as we can see above. The quadratic terms cross on graviton $h_{\mu\nu}$
and gauge field $a_{\mu}$ are,
\begin{eqnarray}
 &  & -\frac{1}{2}a\Gamma_{Ag}^{A}\kappa hS_{,A}=-\frac{1}{2}\int d^{4}xd^{4}x'd^{4}x''a_{\nu}(x')\Gamma_{A_{\nu}(x')g_{\alpha\beta}(x'')}^{A_{\mu}(x)}\kappa h_{\alpha\beta}(x'')\left.\frac{\delta S}{\delta A_{\mu}(x)}\right|_{\bar{\varphi}^{i}}\nonumber \\
 & = & -\frac{\kappa}{8}\int d^{4}x\left[\delta^{\alpha\beta}\delta_{\mu}^{\nu}-2\delta^{\nu(\alpha}\delta_{\mu}^{\beta)}\right]\partial_{\lambda}\bar{F}{}^{\mu\lambda}h_{\alpha\beta}a_{\nu}
\end{eqnarray}
which involves the term $\partial_{\lambda}\bar{F}{}^{\mu\lambda}$ and
will not contribute the corrections to $F^{2}$ operator in our calculation.
The quadratic terms on graviton $h_{\mu\nu}$ are given by,
\begin{eqnarray}
 &  & -\frac{1}{2}\kappa h\Gamma_{gg}^{g}\kappa hS_{,g}=-\frac{\kappa^{2}}{2}\int d^{4}xd^{4}x'd^{4}x''h_{\mu\nu}(x')\Gamma_{g_{\mu\nu}(x')g_{\rho\sigma}(x'')}^{g_{\lambda\tau}(x)}h_{\rho\sigma}(x'')\left.\frac{\delta S}{\delta g_{\lambda\tau}(x)}\right|_{\bar{\varphi}^{i}}\nonumber \\
 & = & \frac{\kappa^{2}}{4}\int d^{4}xh_{\mu\nu}h_{\rho\sigma}\left[\frac{1}{4}\delta^{\mu\rho}\delta^{\sigma\nu}\bar{F}{}^{2}-\frac{1}{8}\delta^{\mu\nu}\delta^{\rho\sigma}\bar{F}{}^{2}+\frac{1}{2}\delta^{\rho\sigma}\bar{F}{}^{\mu}{}_{\alpha}\bar{F}{}^{\nu\alpha}-\delta^{\nu\rho}\bar{F}{}^{\mu}{}_{\alpha}\bar{F}{}^{\sigma\alpha}\right]
\end{eqnarray}
 These are interacting terms between graviton and gauge boson from
the connection terms, which is crucial for gauge condition independent
calculation.

\subsection{Gauge Fixing terms $\frac{1}{4\Omega}\eta^{i}K_{\alpha i}K_{j}^{\alpha}\eta^{j}$}

For getting proper propagators, we shall also include the gauge fixing
term. The Landau-DeWitt gauge is defined by eq. (\ref{eq:LDgauge}) with
$\chi_{\alpha}=K_{\alpha}^{i}[\bar{\varphi}]g_{ij}[\bar{\varphi}]\eta^{j}=0$. For the gravity, we have
\begin{eqnarray}
\chi_{\alpha} & = & \int d^{4}x'd^{4}x''\left(\left[-\delta_{\mu\alpha}\partial'_{\nu}-\delta_{\alpha\nu}\partial'_{\mu}\right]\delta(x',x)\frac{1}{\kappa^{2}}\left(\delta^{\mu(\rho}\delta^{\sigma)\nu}-\frac{1}{2}\delta^{\mu\nu}\delta^{\rho\sigma}\right)\delta(x',x'')\kappa h_{\rho\sigma}(x'')\right.\nonumber \\
 &  & +\left[-\bar{A}_{\mu,\alpha'}(x')-\bar{A}_{\alpha}(x')\partial'_{\mu}\right]\delta(x',x)\delta^{\mu\nu}\delta(x',x'')a_{\nu}(x'')\biggl)\nonumber \\
 & = & \frac{2}{\kappa}\left[\partial^{\mu}h_{\mu\alpha}-\frac{1}{2}\partial_{\alpha}h\right]+\left[a^{\mu}\bar{F}_{\mu\alpha}+\bar{A}_{\alpha}\partial^{\mu}a_{\mu}\right]
\end{eqnarray}
with $h=\delta_{\mu\nu}h^{\mu\nu}$. For the gauge field, we yield
\begin{eqnarray}
\chi & = & \int d^{4}x'd^{4}x''\left[\partial'_{\mu}\delta(x',x)\right]\delta^{\mu\nu}\delta(x',x'')a_{\nu}(x'')=-\partial^{\mu}a_{\mu}
\end{eqnarray}
Thus the Landau-DeWitt gauge conditions ($\omega=1$) are found to be
\begin{eqnarray}
\chi_{\lambda} & = & \frac{2}{\kappa}(\partial^{\mu}h_{\mu\lambda}-\frac{1}{2}\partial_{\lambda}h)+\omega(\bar{A}_{\lambda}\partial^{\mu}a_{\mu}+a^{\mu}\bar{F}_{\mu\lambda})\label{eq:GaugeForGraviton}\\
\chi & = & -\partial^{\mu}a_{\mu}\label{Eq:GaugeForGauge}
\end{eqnarray}
 where $\omega$ is a parameter introduced \cite{DJToms} for a comparison
with the traditional background-field method with harmonic gauge($\omega=0$).
It is tempting to impose $\partial^{\mu}a_{\mu}=0$ in eq.(\ref{eq:GaugeForGraviton}),
we shall discuss it late on as this term can bring
quartic divergences which may break the $U(1)$ gauge invariance.

The gauge fixing term can be written explicitly as
\begin{equation}
S_{GF}=\frac{1}{4\Omega}\eta^{i}K_{\alpha i}K_{j}^{\alpha}\eta^{j}=\frac{1}{4\xi}(\chi_{\lambda})^{2}+\frac{1}{4\zeta}(\chi)^{2}
\end{equation}
 where $\xi$ and $\zeta$ are gauge fixing parameters for gravity
and gauge fields, respectively. The gauge fixing terms are given by
\begin{eqnarray}
S_{GF} & = & \frac{1}{\kappa^{2}\xi}\left[\partial^{\mu}h_{\mu\lambda}-\frac{1}{2}\partial_{\lambda}h\right]^{2}+\frac{1}{4\zeta}\left[\partial^{\mu}a_{\mu}\right]^{2}+\frac{\omega^{2}}{4\xi}\left[\bar{A}_{\lambda}\partial^{\mu}a_{\mu}+a^{\mu}\bar{F}_{\mu\lambda}\right]^{2}\nonumber \\
 &  & +\frac{\omega}{\kappa\xi}\left[\partial^{\mu}h_{\mu\lambda}-\frac{1}{2}\partial_{\lambda}h\right]\left[\bar{A}_{\lambda}\partial^{\mu}a_{\mu}+a^{\mu}\bar{F}_{\mu\lambda}\right]
\end{eqnarray}

\subsection{Ghost part}

It has to include the ghost's contributions as we are working
in a gauge that induces the ghost-gauge coupling. The action of the ghost
part is
\begin{eqnarray*}
S_{GH} & = & \bar{c}_{\alpha}Q^{\alpha}{}_{\beta}[\bar{\varphi}]c^{\beta}=\bar{c}^{\lambda}\frac{\delta\chi_{\lambda}}{\delta\varphi^{i}}\frac{\delta\varphi^{i}}{\delta\epsilon^{\rho}}\biggl|_{\bar{\varphi}^{i}}c^{\rho}+\bar{c}^{\lambda}\frac{\delta\chi_{\lambda}}{\delta\varphi^{i}}\frac{\delta\varphi^{i}}{\delta\epsilon}\biggl|_{\bar{\varphi}^{i}}c+\bar{c}\frac{\delta\chi}{\delta\varphi^{i}}\frac{\delta\varphi^{i}}{\delta\epsilon}\biggl|_{\bar{\varphi}^{i}}c+\bar{c}\frac{\delta\chi}{\delta\varphi^{i}}\frac{\delta\varphi^{i}}{\delta\epsilon^{\rho}}\biggl|_{\bar{\varphi}^{i}}c^{\rho}\\
 & = & S_{GH0}+S_{GH1}+S_{GH2}
\end{eqnarray*}
where $c^{\rho}$ and $c$ denotes the corresponding ghost for gravity
and gauge feilds, respectively. For the evaluation of one-loop two-point Green's function,
we can drop terms with quantum fields $h_{\mu\nu}\text{ and }a_{\mu}$
since the ghost and its anti-ghost will form a loop, one more quantum
field needs contract with another quantum field, forming the second
loop. The action is the sum of the following four parts,
\begin{eqnarray*}
 &  & \bar{c}\frac{\delta\chi}{\delta\varphi^{i}}\frac{\delta\varphi^{i}}{\delta\epsilon}\biggl|_{\bar{\varphi}^{i}}c=\int d^{4}xd^{4}x'd^{4}\bar{x}\left[\bar{c}(x)\frac{\delta\chi(x)}{\delta\varphi^{i}(x')}K^{\varphi^{i}}(x',\bar{x})c(\bar{x})\right]\\
 & = & \int d^{4}x\left[-\bar{c}(x)\right]d^{4}x'\left[\partial^{\mu}\delta(x,x')\partial_{\mu}^{'}c(x')\right]=\int d^{4}x\left[-\bar{c}\square c\right]
\end{eqnarray*}
which is the free part of Lagrangian for the ghost of the $U(1)$ gauge theory. And
\begin{eqnarray*}
 &  & \bar{c}^{\lambda}\frac{\delta\chi_{\lambda}}{\delta\varphi^{i}}\frac{\delta\varphi^{i}}{\delta\epsilon^{\rho}}\biggl|_{\bar{\varphi}^{i}}c^{\rho}=\int d^{4}xd^{4}x'd^{4}\bar{x}\left[\bar{c}^{\lambda}(x)\frac{\delta\chi_{\lambda}(x)}{\delta\varphi^{i}(x')}K^{\bar{\varphi}^{i}}{}_{\rho}(x',\bar{x})c^{\rho}(\bar{x})\right]\\
 & = & \int d^{4}x\Biggl[-\frac{2}{\kappa^{2}}\bar{c}^{\lambda}\square c_{\rho}-\omega\bar{c}^{\lambda}\bar{F}_{\;\lambda}^{\nu}\bar{A}_{\nu,\rho}c^{\rho}-\omega\bar{c}^{\lambda}\bar{F}_{\;\lambda}^{\nu}\bar{A}_{\rho}\partial_{\nu}c^{\rho}-\omega\bar{c}^{\lambda}\bar{A}_{\lambda}\bar{A}_{\rho}\square c^{\rho}-\omega\bar{c}^{\lambda}\bar{A}_{\lambda}\bar{A}_{\nu,\rho}\partial^{\nu}c^{\rho}\Biggl]
\end{eqnarray*}
where the first term in the bracket gives the free part of Lagrangian for the
gravitational ghost introduced by fixing the general coordinate transformation,
and other terms are treated as interactions. These interactions are
proportional to $\omega$, and indicate that in harmonic gauge $\omega=0$,
gravitational ghost and anti-ghost do not interact with gauge fields. However,
there are still interactions like $\bar{c}Ac^{\mu}$ in the Lagrangian
for arbitrary $\omega$.
\begin{eqnarray*}
 &  & \bar{c}^{\lambda}\frac{\delta\chi_{\lambda}}{\delta\varphi^{i}}\frac{\delta\varphi^{i}}{\delta\epsilon}\biggl|_{\bar{\varphi}^{i}}c=\int d^{4}xd^{4}x'd^{4}\bar{x}\left[\bar{c}^{\lambda}(x)\frac{\delta\chi_{\lambda}(x)}{\delta\varphi^{i}(x')}K^{\bar{\varphi}^{i}}(x',\bar{x})c(\bar{x})\right]\\
 & = & \omega\int d^{4}x\left[\bar{c}^{\lambda}\bar{A}_{\lambda}\square c+\bar{c}^{\lambda}\bar{F}_{\;\lambda}^{\mu}\partial_{\mu}c\right],
\end{eqnarray*}
 and
\begin{eqnarray*}
 &  & \bar{c}\frac{\delta\chi}{\delta\varphi^{i}}\frac{\delta\varphi^{i}}{\delta\epsilon^{\rho}}\biggl|_{\bar{\varphi}^{i}}c^{\rho}=\int d^{4}xd^{4}x'd^{4}\bar{x}\left[\bar{c}(x)\frac{\delta\chi(x)}{\delta\varphi^{i}(x')}K^{\bar{\varphi}^{i}}{}_{\rho}(x',\bar{x})c^{\rho}(\bar{x})\right]\\
 & = & \int d^{4}x\left[\bar{c}\bar{A}_{\:,\mu\rho}^{\mu}c^{\rho}+\bar{c}\bar{A}_{\mu,\rho}c^{\rho,\mu}+\bar{c}\bar{A}_{\rho}\square c^{\rho}+\bar{c}\bar{A}_{\rho,\mu}c^{\rho,\mu}\right].
\end{eqnarray*}

\newpage


\end{document}